\documentclass[amsmath,amssymb,twocolumn]{revtex4}
\usepackage[dvips]{graphicx}
\usepackage{bm}

\begin{document}

\title{Full counting statistics for transport through a molecular quantum dot magnet}

\author{K.-I. Imura$^1$, Y. Utsumi$^1$, and T. Martin$^{1,2,3}$}
\affiliation{
${}^{1}$~Condensed Matter Theory Laboratory, RIKEN, Wako, Saitama 351-0198, Japan\\
${}^2$~Centre de Physique Th\'eorique, Case 907 Luminy, F-13288 Marseille Cedex 9, France \\
${}^3$~Universit\'e de la M\'edit\'erann\'ee, F-13288 Marseille Cedex 9, France}

\begin{abstract}
Full counting statistics (FCS) for the transport through 
a molecular quantum dot magnet is studied theoretically
in the incoherent tunneling regime.
We consider a model describing a single-level quantum dot, 
magnetically coupled to an additional local spin, 
the latter representing the total molecular spin $s$. 
We also assume that the system is in the strong Coulomb blockade regime,
i.e., double occupancy on the dot is forbidden.
The master equation approach to FCS introduced in Ref. \onlinecite{bagrets}
is applied to derive a generating function yielding the FCS of charge and current.
In the master equation approach, 
Clebsch-Gordan coefficients appear in the transition probabilities,
whereas the derivation of generating function
reduces to solving the eigenvalue
problem of a modified master equation with counting fields.
To be more specific, one needs only the eigenstate which collapses smoothly to
the zero-eigenvalue stationary state in the limit of vanishing counting fields.
We discovered that in our problem with arbitrary spin $s$,
some quartic relations among Clebsch-Gordan coefficients allow us
to identify the desired eigenspace without solving the whole problem.
Thus we find analytically the FCS generating function in the following two cases:
i) both spin sectors ($j=s\pm 1/2$) lying in the bias window, 
ii) only one of such spin sectors lying in the bias window. 
Based on the obtained analytic expressions, we also developed a numerical analysis in order 
to perform a similar contour-plot of the joint charge-current distribution function, 
which have recently been introduced in Ref. \onlinecite{utsumi_condmat},
here in the case of molecular quantum dot magnet problem.
\end{abstract}

\maketitle

\section{Introduction} 

Molecular electronics has emerged as one of the promising subfields of nanophysics.
Individual nano objects such as molecules and carbon nanotubes can be connected to 
leads of different nature, and the degrees of freedom of molecules could be used in 
principle to modulate quantum transport between the leads
\cite{nitzan,book,tsukada}: 
it is hoped that these degrees of freedom could allow specific functions of nanoelectronics.
A large body of work has focused on the potential applications of the specific energy 
spectrum of molecules, as well as their vibrational degrees of freedom.
\cite{some_reviews}
Recently, there has been some focus on transport through molecules which have their 
own spin, for instance because a specific atom of this molecular system possesses 
a magnetic moment. This is unlike ``usual'' molecular quantum dots where the molecular 
spin corresponds to the effective spin of the itinerant electron, giving rise for example 
to Kondo physics.
\cite{utsumi_martinek}
There are many experimental proposals for such molecular magnets.
Mn$_{12}$ acetates or carbon fullerenes with one or more atoms trapped inside are examples.
Depending on the sample, and on the structure of the molecule, the spin orientation 
can have a preferred axis  or it can have a more isotropic nature.
There have been recent experiments on similar systems, with applications to spin valves. 
\cite{heersche,romeike_incoh}
In this paper we focus on the isotropic regime.
As in Ref. \cite{elste_timm}, where the current through an endohedral nitrogen doped 
fullerene was computed, we describe transport in the incoherent tunneling regime, 
and proceed to compute the full counting statistics for transport 
through a molecular quantum dot 
where the dot electrons have an exchange coupling with an impurity spin.

Indeed, in mesoscopic physics, \cite{nazarov,fazio}
the current voltage characteristics alone is in general 
not sufficient to fully characterize transport.
The current fluctuations can provide additional information on the correlations in 
molecular quantum dot, but so do the higher current moments, or cumulants, 
such as the ``skewness'' (third order) and the ``sharpness'' (fourth order).
Such higher order cumulants add up to the so-called full counting statistics (FCS).
\cite{levitov}
Yet here we are interested in obtaining the FCS of such devices using the master 
equation approach. \cite{bagrets}
We will also take the opportunity to present in detail the general framework for 
studying joint distributions of electron occupation of the dot
and of current. A recent work by one of the authors
\cite{utsumi_condmat} 
applied to a single level quantum dot shows that there are 
clear correlations between current and occupancy of the dot: 
there, the quantum fluctuations caused by increasing the tunnel coupling to the leads 
has a definite impact on the number distribution.  
Here we will extend these concepts to the case of a molecular quantum dot with an impurity spin.   

The FCS has remained until recently a highly theoretical concept, 
with few experimental applications. 
The third moment has been measured for a tunnel junction,
\cite{reulet,reznikov}
as the effects of the measuring apparatus have been pointed out. 
\cite{kindermann}
For the regime of incoherent transport through quantum dots two 
outstanding experiments have been performed:
\cite{fujisawa,gustavsson_ensslin} 
a point contact is placed in the close vicinity of the quantum dot, 
and by monitoring the current in this point contact, 
it is possible to obtain information on the time series of the current in the dot. 
From this time series, higher moments of the current can be computed.
These recent experiments thus provide additional motivation to study the FCS in this novel
system of a molecular quantum dot with an external spin.      

This paper thus aims at calculating the FCS of transport through a molecular
quantum dot magnet, under the following circumstances:
\begin{enumerate}
\item The quantum dot is in the strong Coulomb blockade regime, i.e., the system is
subject to strong electronic correlation.
\item The coupling between the dot and the reservoirs is relatively weak, and tunneling of
electrons between them can be considered incoherent (incoherent tunneling regime).
\item The molecular spin is isotropic, and coupled to the spin of conduction 
electron through
a simple exchange interaction.
\end{enumerate}
Assumption (2) allows us to employ the master equation approach for
studying the transport through molecular quantum dot.
As assumption (1) forbids double occupancy on the dot, we consider 
the cases where the dot is either empty, or occupied by only one conduction 
electron.

When the dot is empty, the molecular spin $s$, together with its $z$-component
$s_z$, is naturally a good quantum number,
whereas if the dot is occupied by one conduction electron, then it becomes
the total angular momentum $j=s+\sigma$ which is a good quantum number on the dot. 
Let us now follow the time evolution of the dot state, e.g., using a master equation.
The number $n$ of conduction electrons on the dot fluctuates in time
between $n=0$ and $n=1$.
Such transitions are controlled by the so-called rate matrix $L$,
which will be more carefully defined in Sec. II.
This rate matrix $L$ can be decomposed into $2\times 2=4$ charge blocks ,
corresponding to two charge sectors $n=0,1$. 
Then each block is further characterized by either the molecular
spin $s_z$ or the total angular momentum $j$ (and $j_z$) depending on whether
$n=0$ or $n=1$. 
Naturally, off-diagonal blocks of the rate matrix describe
transition between different angular momentum states.
The rate of such transitions is proportional to the square of a Clebsch-Gordan coefficient
between the initial and the final angular momentum states.

In the master equation approach to FCS, the derivation of 
FCS generating function reduces to solving the eigenvalue problem of 
a modified master equation with counting fields.
By using an algebraic identity, this eigenvalue problem can be projected down
to a slightly more complicated problem within the $n=0$ charge sector,
which can be fully characterized by the molecular spin $s_z$.
Our main discovery is that some identities among Clebsch-Gordan coefficients
allow us to identify the desired eigenspace without solving the whole problem.
In this reduced problem Clebsch-Gordan coefficients appear at quartic order,
since off-diagonal blocks are projected, together with the $n=1$ charge sector, 
down to the $n=0$ subspace.
Based on the analytic expressions for the FCS generating function thus obtained, 
we also developed numerical analysis to perform a contour-plot of 
the joint charge-current distribution function.

The paper is organized as follows. 
The first two sections after Introduction are devoted to the preparatory stage, 
i.e., Section II describes our model for the isotropic molecular quantum dot magnet, 
while Section III reviews the master equation approach to FCS. 
Then, in Section IV the analytic expressions for the FCS generating function will be 
obtained, with the help of quartic identities among Clebsch-Gordan coefficients.
Section V describes the contour plot of joint distribution functions, before the
paper is concluded in Section VI.

\section{Molecular quantum dot magnet}

We consider a molecule connected to two electrodes,
which are specified by an index $\chi=\pm 1$.
$\chi=1$ corresponds, e.g., to the left ($\chi=-1$ to the right)
electrode.

We focus on the strong Coulomb blockade regime, assuming that the number
of electrons which can be added to the dot is restricted to one. 
Because the extra-charge electron state on the dot carries a spin, 
it is locally coupled to the impurity spin on the molecule
via an exchange coupling Hamiltonian.  

\subsection{Model Hamiltonian}

The total Hamiltonian $H$ of the system consists of two parts: 
(i) the Hamiltonian $H_0$ of the isolated parts, and 
(ii) the tunneling Hamiltonian $H_{\rm tun}$, i.e.,
$H=H_0+H_{\rm tun}$.
$H_0$ can be decomposed further into the dot and lead parts: 
$H_0=\sum_{\chi=\pm 1}H_{\rm lead}^{(\chi)}+H_{\rm dot}$.
For normal metal leads, the Hamiltonian of left ($\chi=1$) and right ($\chi=-1$) 
leads reads,
$H_{\rm lead}^{(\chi)}=
\sum_{k\sigma} \epsilon_{k}^{(\chi)}c_{k\sigma}^{(\chi)\dagger}c_{k\sigma}^{(\chi)}$,
where $c_{k\sigma}^{(\chi)}$ annihilates  
an electron with momentum $k$, spin $\sigma$ and energy $\epsilon_{k}^{(\chi)}$ 
in the lead $\chi$.
The tunnel Hamiltonian reads:
$H_{\rm tun}=\sum_{k\sigma\chi} T_{k\sigma}^{(\chi)} c_{k\sigma}^{(\chi)\dagger}d_\sigma 
+ {\mathrm h.c.}$,
where the operator $d$ ($d^\dagger$) annihilates (creates) an electron  
on the single dot level.
The dot Hamiltonian reads:
\begin{equation}
H_{\rm dot}= \epsilon_{\rm dot} \sum_{\sigma} n_{\sigma} + U n_{\uparrow}n_{\downarrow} 
+ H_{\rm spin},
\end{equation}
where
$n_{\sigma}=d^\dagger_{\sigma}d_{\sigma}$ is the number of electrons on the dot level
with spin $\sigma$, and
$U$ is a charging energy contribution related to the capacitance with respect to the leads 
and to the gate. 
It will be assumed to be large in the present work, in order to prohibit double 
occupancy.
Finally, $H_{\rm spin}$ is the coupling of the dot electron spin $\vec{\sigma}$
with the impurity spin $s$. This is finite, by definition, only when there is
an electron on the dot:
\begin{equation}
H_{\rm spin}=\left\{
\begin{array}{l}
0\ \ \ (n=0)
\\
g \vec{\sigma}\cdot\vec{s}\ \ \ (n=1)
\end{array}
\right.,
\end{equation}
where $g$ is the exchange coupling constant.
As expected the eigenstates of this contribution to the Hamiltonian are the 
total angular momentum states of 
$\vec{j} \equiv \vec{\sigma}+\vec{s}$. 
This is seen by rewriting the exchange term as 
$\vec{\sigma}\cdot\vec{s}=(j^2-s^2-\sigma^2)/2$.
The $n=0$ sector is characterized simply by the spin $s$ of the
magnetic impurity, and has $2s+1$ degenerate states,
$|n=0,s,s_z\rangle$.
When $n=1$, the coupling $g$ between the impurity and the dot  
electron spins requires to diagonalize the interaction with
the total angular momentum states, i.e.,
$|n=1,j=s\pm 1/2,j_z\rangle$.
The two spin subsectors, $j=s+1/2$ and $j=s-1/2$
correspond to an energy eigenvalue of, respectively, 
$\epsilon_{s+1/2}=\epsilon_{\rm dot} + g s/2$ and 
$\epsilon_{s-1/2}=\epsilon_{\rm dot} - g (s+1)/2$.
The two subsectors each have a degeneracy of  
$2s+2$ and $2s$, respectively.

In our model Hamiltonian,
the charging effects are taken into account via the bias-voltage dependence of
the position of the  dot level $\epsilon_{\rm dot}$ 
with respect to the chemical potentials of the leads, $\mu_{L,R}$. 
A gate voltage $V_g$ and gate capacitance $C_g$ can be included 
in order to move the dot levels up and down. 
Here we focus on the case where the bias voltage
always dominates on the temperature, so for later purpose it is sufficient to 
specify which level and how many of these are included in the bias window.

\subsection{Incoherent tunneling regime - the master equation}

The electron dynamics of transport through a molecular quantum dot
in the incoherent tunneling regime can be described by a master equation.
\cite{kampen}
In such systems as our molecular quantum dot Hamiltonian,
the master equation describes a stochastic evolution 
in the space spanned by the eigenstates of the non-perturbed Hamiltonian $H_0$.
Such states are, of course, of quantum mechanical nature, but the evolution of the
system is governed by a deterministic stochastic process.
Jumps between different states are to be stochastic events, 
and are also to be Markovian (no correlation between successive tunneling events).
Such a situation is justified in the so-called incoherent tunneling regime,
where the the non-perturbed Hamiltonian $H_0$ is only weakly
perturbed by the tunneling Hamiltonian $H_{\rm tun}$, 
and the tunneling of electrons through the dot is still considered to be
sequential.

A simplification specific to our model is that the lead degrees
of freedom can be traced out, and the master equation describes 
actually the time evolution of the dot states.
We will often denote a (quantum mechanical) dot state $|n,j,j_z\rangle$, 
simply as $\alpha$ in the master equation. 
The probability $p_\alpha (t)$ with which the system, i.e., 
our molecular quantum dot, finds itself in a state $\alpha$ at time $t$ 
satisfies,
\begin{equation}
{d p_\alpha (t) \over dt}
=\gamma_\alpha p_\alpha (t)
+\sum_\beta \Gamma_{\alpha\beta}p_\beta (t).
\label{mas}
\end{equation}
$\Gamma_{\alpha\beta}$ is the transition probability for a jump, 
$\alpha\leftarrow\beta$, which can be calculated, 
e.g., from Fermi's golden rule.
The subscripts $\alpha$ and $\beta$ take all the possible values,
$\alpha_1,\alpha_2,\cdots$
in the space of physically relevant states.
Conservation of probability imposes that the sum of all the
components {\it in a given column} $\beta$ of matrix $L$
vanishes.
Thus $\gamma_\alpha=-\sum_\beta\Gamma_{\beta\alpha}$ denotes
the decay rate of a state $\alpha$.
In matrix notation, this reads
$dp(t)/dt=Lp(t)$, where $L=\gamma+\Gamma$,
is called the rate matrix.
$\gamma$ and $\Gamma$ are, respectively, the diagonal and off-diagonal
parts of the rate matrix $L$.

In order to apply Fermi's golden rule for evaluating $\Gamma_{\alpha\beta}$, 
one must go back to the Hamiltonian, 
$H_0$ 
of the isolated parts, and one consider the transition rate 
$\Gamma_{fi}$ associated with a tunneling perturbation $H_{\rm tun}$,
from an initial state $|i\rangle$ to a final state $|f\rangle$:
\begin{equation}
\Gamma_{fi}={2\pi\over\hbar}
\big|\langle i|H_{\rm tun}|f\rangle\big|^2
\delta(E_i-E_f)~..
\label{gold}
\end{equation}
The eigenstates $|i\rangle,|f\rangle$
are many electron states, but they factorize into a direct product of 
many-electron lead states and a dot state.
Therefore, after taking into account all the lead degrees of freedom,
the above transition rates are rewritten in the form of
transition rates between different dot states $\alpha,\beta$,
i.e., $\Gamma_{\alpha\beta}$ for a transition, $\alpha\leftarrow\beta$.

We also make the standard assumption that the hopping amplitude 
$T_{k\sigma}^{(\chi)}$
depends weakly on energy for states close to the Fermi levels of the leads.
For normal leads it is also spin independent.
Thus Fermi's golden rule gives basically a bare tunneling rates, defined as 
$\Gamma_{L,R}=2\pi \nu T_{L,R}^2$,
weighted by a Clebsch-Gordan coefficient squared,
which we will discuss later.
$\nu$ is the (constant) density of states at the Fermi level in the leads.
We also use the explicit notation $L,R$ rather than $\chi$ for
specifying the leads.

Before giving explicit matrix elements of the rate matrix $L$,
it is convenient to decompose it into four blocks, corresponding to the 
two charge sectors $n=0,1$ as
\begin{equation}
L=
\left(\begin{array}{cc}
L_{00} & L_{01} \\
L_{10} & L_{11}
\end{array}\right)
\equiv
\left(\begin{array}{cc}
A & B \\
C & D
\end{array}\right).
\label{ratemat}
\end{equation}
The off-diagonal block matrices $B$ and $C$ are rectangular matrices
of size, (i) $(2s+1)\times (4s+2)$ and (ii) $(4s+2)\times (2s+1)$,
respectively.
They describe tunneling of an electron either 
(i) out of the dot into one of the reservoirs, or
(ii) onto the dot from one of the reservoirs.
Both have two spin sectors,
corresponding to two possible values of the total angular momentum
$j=s\pm 1/2$, i.e., $B=(B_{s-1/2},B_{s+1/2})$ and $C=(C_{s-1/2},C_{s+1/2})^T$,
where the $C^T$ is, for example, a transposed matrix of $C$.
The $(s_z,j_z)$-component of the submatrices $B_{s\pm1/2}$ and
the $(j_z,s_z)$-component of $C_{s\pm1/2}$ are given, respectively, as
\begin{eqnarray}
&&B_j(s_z,j_z)=
\nonumber \\
&&\big[\Gamma_L\{1-f(\epsilon_j-\mu_L)\}+\Gamma_R\{1-f(\epsilon_j-\mu_R)\}\big]
\nonumber \\ 
&&\times|\langle s,s_z;1/2,\sigma_z|j,j_z\rangle|^2
\label{b}\\
&&C_j(j_z,s_z)=
\nonumber \\
&&\big[\Gamma_L f(\epsilon_j-\mu_L)+\Gamma_R f(\epsilon_j-\mu_R)\big]
\nonumber \\
&&\times|\langle s,s_z;1/2,\sigma_z|j,j_z\rangle|^2,
\label{c}
\end{eqnarray}
where
$j=s\pm 1/2$, $j_z=-j,\cdots,j$.

The diagonal blocks $A$ and $D$ are themselves diagonal matrices,
because the dot Hamiltonian is diagonal in the chosen basis. Their elements
are specified by recalling that the sum of all the components in a given column
of $L$ vanishes identically, as a result of the conservation
of probability.

The matrix $D$ has two spin sectors,
$D={\rm diag}(D_{s-1/2},D_{s+1/2})$.
The two diagonal blocks $D_j$ are also diagonal matrices,
$D_j(j_z,j'_z)=D_j(j_z,j_z)\delta (j_z,j'_z)$.
Let us focus on the column of $L$ belonging to the sector $j$ and $j_z$.
The conservation of probability associated with this column 
requires,
\begin{eqnarray}
D_j(j_z,j_z)&=&-\sum_{s_z=-s,\cdots,s}B_j(s_z,j_z)\nonumber \\
&=&-\Gamma_L\big[1-f(\epsilon_j-\mu_L)\big]
-\Gamma_R\big[1-f(\epsilon_j-\mu_R)\big]~,\nonumber\\
\label{d}
\end{eqnarray}
i.e., $D_j(j_z,j_z)$ actually does not depend on $j_z$: 
this simplification is due to the identity (\ref{jzquad}).
Similar quadratic relations among Clebsch-Gordan coefficients are
also summarized in the Appendix.

The diagonal elements of matrix $A$ can be determined in the same manner.
Recall that $A(s_z,s'_z)=A(s_z,s_z)\delta(s_z,s'_z)$.
This time one fixes $s_z$ in order to specify a particular column.
Then, in order to apply the conservation law, 
we sum over the two spin sectors $j$ and their associated $j_z$,
to find,
\begin{eqnarray}
&&A(s_z,s_z)=-\sum_{j=s\pm 1/2}\sum_{j_z=-j,\cdots,j}C_j(j_z,s_z)
\nonumber \\
&&=-{2s\over 2s+1}
\Big[\Gamma_L f(\epsilon_{s-1/2}-\mu_L)+\Gamma_R f(\epsilon_{s-1/2}-\mu_R)\Big]
\nonumber \\
&&-{2s+2\over 2s+1}
\Big[\Gamma_L f(\epsilon_{s+1/2}-\mu_L)+\Gamma_R f(\epsilon_{s+1/2}-\mu_R)\Big].
\label{a}
\end{eqnarray}
Here we used the identity (\ref{szquad}) in the Appendix.

The explicit matrix elements of the rate matrix are thus given.
The discussion now turns to the description of FCS in the framework of master equation.


\section{Master equation approach to FCS}

We are interested in the statistical correlations of such observables
as charge $N(t)$ or current $I(t)$,
measured during a time interval $t_0<t<t_0+\tau$.
In the master equation approach, an observable $Q_f(t)$ 
($f=1,2,\cdots; Q_1(t)=N(t),Q_2(t)=I(t),\cdots$),
is introduced as a stochastic variable which evolves as a function of time
through the master equation.
On the other hand, what is experimentally relevant is
their time average during the measurement time $\tau$,
\begin{equation}
Q_f={1\over\tau}\int_{t_0}^{t_0+\tau} dt Q_f(t).
\label{average}
\end{equation}
The quantity of our eventual interest is the joint statistical 
distribution function $P(q_1,q_2,\cdots)$ of such observables
averaged during time $\tau$, i.e.,
\begin{equation}
P(q_1,q_2,\cdots)=\big\langle
\delta(q_1-Q_1)\delta(q_2-Q_2)\cdots
\big\rangle,
\label{jdf}
\end{equation}
where $\langle\cdots\rangle$ represents a stochastic average,
whose meaning will be more rigorously defined in Sec. III B.
The joint distribution function $P(q_1,q_2,\cdots)$ is related to
the generating function via a Fourier transformation:
\begin{eqnarray}
\sum_{q_1,q_2,\cdots} P(q_1,q_2,\cdots)\exp\Big[i\sum_{f=1,2,\cdots}q_f\Phi_f\Big]
\nonumber \\
=\exp\big[\Omega(\Phi_1,\Phi_2,\cdots)\big],
\nonumber
\label{cgf}
\end{eqnarray}
where $\Omega(\Phi_1,\Phi_2,\cdots)$ 
is called the cumulant generating function (CGF).
The ($n_1,n_2,\cdots$)-th order cumulant $C_{n_1,n_2,\cdots}$ 
is deduced from the generating function, $\Omega(\Phi_1,\Phi_2,\cdots)$
as,
\begin{equation}
C_{n_1,n_2,\cdots}=(-i)^n 
{\partial^{n_1+n_2+\cdots}\over\partial\Phi_1^{n_1}\partial\Phi_2^{n_2}\cdots}
\Omega(\Phi_1,\Phi_2,\cdots)\Big|_{\Phi_1,\Phi_2,\cdots=0}.
\nonumber
\end{equation}
Note that the CGF, $\Omega(\Phi_1,\Phi_2,\cdots)$ and 
the full distribution function $P(q_1,q_2,\cdots)$ 
contain the same information.
In the following, 
we will focus on how to calculate such CGF in the framework of the
master equation approach.

The purpose of this section is to review the master equation
approach to FCS developed in Ref. \onlinecite{bagrets},
and adapt it to a more specific case of our interest.
A close comparison with the time-dependent perturbation
theory of quantum mechanics 
reduces the calculation of CGF to solving 
a modified master equation with counting fields.

\subsection{Two types of observables and their counting statistics}

In Secs. IV and V we discuss counting statistics 
of charge $N$ and current $I$.
Other observables, such as the total angular momentum 
$J$ on the dot, its $z$-component $J_z$, or the 
spin current $I_s$ through the dot, can be handled 
in a similar way, which we outline below.
Note that we can make a distinction between two category of 
observables: ``charge-like'' and ``current-like'' observables.
The charge $N$ (or $J,\cdots$) is a property of the state of the dot, 
whereas the charge current, $I_c$ (or $I_s,\cdots$) is 
associated with transitions between different occupational states.
The former has eigenvalues of a quantum mechanical
operator that can be simultaneously diagonalized with $H_0$, 
whereas the latter is related to the nature of the transitions.
We may also classify the counting fields $\Phi_1,\Phi_2,\cdots$
into two categories: 
$\xi_1,\xi_2,\cdots$ and 
$\eta_1,\eta_2,\cdots$.

In order to give an unambiguous meaning to the stochastic average
$\langle\cdots\rangle$ 
in Eq. (\ref{cgf}) or (\ref{jdf}) in the context of master equation, 
one has to 
go one step {\it backward} in regard to Eq. (\ref{average}),
i.e., one has to follow the time evolution of observables
$Q_f(t)$. One may rewrite Eqs. (\ref{jdf}) and (\ref{cgf}) as
\begin{eqnarray}
&&\exp\Big[\Omega(\xi_1,\xi_2,\cdots; \eta_1,\eta_2,\cdots)\Big]=
\nonumber \\
&&\Big\langle\exp\Big[
i\sum_f \xi_f\int dt N_f(t) 
+i\sum_{f'}\eta_{f'}\int dt I_{f'}(t)
\Big]\Big\rangle~,\nonumber\\
\label{cgf+}
\end{eqnarray}
where the time integrals should be performed, e.g., from $t_0$ to $t_0+\tau$.
The stochastic average can be performed by giving more concrete
expressions to the observables $N_f(t)$ and $I_{f'}(t)$
using the language of stochastic process.

Let us focus on a particular realization of the stochastic process 
by denoting $\{t_j\}$ the sequence of times 
at which jumps between different states occur.
$\{\alpha_j\}$ denotes the sequence of states such that
the system stays in state $\alpha_j$ during the period of
$t_j<t<t_{j+1}$.
A state $|\alpha\rangle$ is specified by the
quantum numbers $n_1,n_2,\cdots$, (e.g.,
$n_1=n$, the charge, $n_2=j$, the total angular momentum,
$n_3=j_z$, its $z$-component, etc...). 
On the other way around, one denotes
by $n_1(\alpha),n_2(\alpha),\cdots$ the quantum numbers for specifying the
state $\alpha$.
With this notation, $N_f(t)$ can be written explicitly
for a Markovian sequence $(\{t_j\}, \{\alpha_j\})$:
\begin{equation}
N_f(t)=\sum_j n_f(\alpha_j)\theta(t-t_j)\theta(t_{j+1}-t),
\label{nf}
\end{equation}
where $\theta(t)$ is the Heaviside function. 

A little more care is needed to give a similar expression to current-like
observables, reflecting the fact that they have a direction.
In order to clarify this point, let us consider the case of our molecular
quantum dot.
The transition of the dot state is caused by tunneling of an electron
onto or out of the dot.
The change of the dot state before and after the transition is completely
specified by the change of the number of electrons $\Delta n$,
and of the total spin $\Delta j$ together with its $z$-component $\Delta j_z$ 
on the dot.
However, in order to define a current associated with the
transition, we need additionally an information on the direction of flow, 
i.e., an information on where the electron goes, to which lead, or from which
lead.
To measure the current, one has to decide also where one measures it.
Such ambiguities are usually removed by considering a {\it net} current
which flows {\it out of} a specific lead, say, $\chi$, i.e.,
by considering $I^{(\chi)}(t)$.
A current $I_f^{(\chi)}=\Delta n_f$ is induced by a tunneling, leading to the 
change of the dot state $\Delta n_f$, such as $\Delta n$, $\Delta j$ and $\Delta j_z$, 
which occurs also through the lead $\chi$.
This suggest in our Markovian sets $(\{t_j\}, \{\alpha_j\})$
to add $\{\chi_j\}$ specifying through which lead the jump at time
$t_j$ occurs.
Once such Markovian sets $(\{t_j\}, \{\alpha_j\},\{\chi_j\})$ 
are given, one can provide explicit formula to current-like observables:
\begin{equation}
I_f^{(\chi)}(t)=\sum_j \Delta n_f (t_j) \delta (\chi-\chi_j) \delta (t-t_j),
\label{if}
\end{equation}
where $\Delta n_f (t_j)=n_f(\alpha_{j+1})-n_f(\alpha_{j})$ specifies the 
nature of transition at time $t=t_j$.

Let us consider the case of our model again.
We separated the rate matrix $L$ into two parts:
the diagonal part $\gamma$ and the off-diagonal part $\Gamma$.
We also introduced a decomposition into four charge sectors, which
can be naturally applied to $\gamma$ and $\Gamma$ as well:
\begin{eqnarray}
\gamma=
\left(\begin{array}{cc}
\gamma_0 & 0 \\
0 & \gamma_1
\end{array}\right)
=
\left(\begin{array}{cc}
L_{00} & 0\\
0 & L_{11}
\end{array}\right),
\nonumber \\
\Gamma=
\left(\begin{array}{cc}
0 & \Gamma_{01} \\
\Gamma_{10} & 0
\end{array}\right)
=
\left(\begin{array}{cc}
0 & L_{01} \\
L_{10} & 0
\end{array}\right).
\nonumber
\end{eqnarray}
Let us focus on the off-diagonal block matrices in $\Gamma$,
$\Gamma_{01}$ and $\Gamma_{10}$.
These two blocks correspond simply to different $\Delta n$,
i.e., $\Gamma_{10}$ to $\Delta n=1$, and $\Gamma_{01}$ to $\Delta n=-1$.
Such submatrices of $\Gamma$ still contain contribution from different leads,
e.g., one can further decompose $\Gamma_{10}$ into
$\Gamma_{10}=\Gamma_{10}^{(L)}+\Gamma_{10}^{(R)}$.
The whole $\Gamma$ can be also decomposed with respect to $\chi$,
i.e., $\Gamma=\sum_\chi\Gamma^{(\chi)}$,
whereas one can divide the matrix $\Gamma$ into smaller subsectors
in terms of different $\Delta n_f$.
Thus very generally one can decompose the transition matrix
$\Gamma$ as
\begin{equation}
\Gamma=\sum_{\chi,\{\Delta n_f\}}\Gamma^{(\chi)}_{\{\Delta n_f\}},
\end{equation}
where 
$\{\Delta n_f\}=(\Delta n,\Delta j,\cdots)$.

\subsection{Master equation with counting fields --- prescription for
obtaining the FCS}

Substituting Eqs. (\ref{nf}) and (\ref{if}) into Eq. (\ref{cgf+})
one finds the expression to be evaluated:
\begin{eqnarray}
\exp\Big[\Omega(\xi_1,\xi_2,\cdots; \eta_1,\eta_2,\cdots)\Big]=
\nonumber \\
\Big\langle\exp\Big[
i\sum_f \xi_f \sum_{j=0}^n (t_{j+1}-t_j) n_f(\alpha_j)+
\nonumber \\
i\sum_{f}\eta_{f}\sum_{j=1}^n \Delta n_f (t_j) \delta (\chi-\chi_j)
\Big]\Big\rangle.
\label{cgf++}
\end{eqnarray}
We need a prescription to compute the stochastic average $\langle\cdots\rangle$.
This is achieved by performing a perturbative expansion of the master equation 
with respect to off-diagonal components $\Gamma$.
The resulting perturbation series is analogous to that of the 
time-dependent perturbation theory of quantum mechanics.
Note that the diagonal part $\gamma$ and the off-diagonal
part $\Gamma$ of the coefficient matrix $L=\gamma+\Gamma$
generally {\it do not commute}.

Because our observables are series of events which occur
successively on the time axis, 
the master equation is rewritten in a way which is 
analogous to the ``interaction representation''
of quantum mechanics,
\begin{eqnarray}
{d\over dt}\tilde{p}(t)&=&\tilde{\Gamma}(t)\tilde{p}(t), 
\nonumber \\
\tilde{p}(t)&=&\exp \big[\gamma(t-t_0)\big] p(t_0),
\nonumber \\
\tilde{\Gamma}(t)&=&e^{-\gamma(t-t_0)}\Gamma e^{\gamma(t-t_0)}.
\label{intrep}
\end{eqnarray}
Developing a perturbative expansion with respect to $\Gamma$, one finds
the analog of the Dyson series for $\tilde{p}(t)$:
\begin{eqnarray}
&&\tilde{p}(t)=p(t_0)+\int_{t_0}^{t} dt' \tilde{\Gamma}(t)\tilde{p}(t')
\nonumber \\
&&=\sum_{n=0}^\infty \int_{t_0<t_1<t_2<\cdots<t_n<t} dt_1dt_2\cdots dt_n 
\nonumber \\
&&~~\times\tilde{\Gamma}(t_n)\cdots \tilde{\Gamma}(t_2)\tilde{\Gamma}(t_1)
p(t_0),
\label{dyson}
\end{eqnarray}
where we used $\tilde{p}(t_0)=p(t_0)$.
Let us focus on the $n$-th order term in $\tilde{p}(t)$.
Because Eq. (\ref{dyson}) is written in matrix notation,
we can further develop it as:
\begin{eqnarray} 
&&\tilde{\Gamma}(t_n)\cdots \tilde{\Gamma}(t_2)\tilde{\Gamma}(t_1)
\Big|_{\alpha_{n}\alpha_{0}}=
\nonumber \\
&&\sum_{\alpha_{n-1},\cdots,\alpha_2,\alpha_1} 
\tilde{\Gamma}_{\alpha\alpha_{n-1}}(t_n)\cdots\tilde{\Gamma}_{\alpha_2\alpha_1}(t_2)
\tilde{\Gamma}_{\alpha_1\alpha_0}(t_1).
\label{jumps}
\end{eqnarray}
One can identify such terms as the realization
of the stochastic process characterized by the sets
$\{t_j\}$ and $\{\alpha_j\}$.
Indeed, each of the $n$-th order term in Eq. (\ref{dyson})
corresponds to a sequence, 
$\alpha_0\rightarrow \alpha_1\rightarrow \alpha_2\rightarrow \cdots$, 
with $n$ jumps located at times $\{t_j\}$.
As the matrix $\gamma$ is the diagonal part of $L$, 
the system remains in the same state $\alpha_j$
with probability $e^{\gamma_{\alpha_0} (t_{j+1}-t_j)}$
between the two jumps $t_{j}<t<t_{j+1}$, whereas the transition
$\alpha_{j-1}\rightarrow \alpha_j$ occurs at time $t_j$.
Thus the probability for a sequence of events 
to occur is found to be,
\begin{eqnarray} 
&&P(\alpha_0\rightarrow \alpha_1\rightarrow \alpha_2\rightarrow \cdots)=
\nonumber \\
&&\cdots
e^{\gamma_{\alpha_2} (t_3-t_2)}
\Gamma_{\alpha_2\alpha_1}
e^{\gamma_{\alpha_1} (t_2-t_1)}
\Gamma_{\alpha_1\alpha_0}
e^{\gamma_{\alpha_0} (t_1-t_0)}
p_{\alpha_0}(t_0).
\nonumber
\end{eqnarray}
The above identification in the framework of Eqs. (\ref{intrep}) and (\ref{dyson})
allows us to give an unambiguous meaning to the stochastic average of
observables in Eq. (\ref{cgf++}).
A prescription for calculating the average is given below.

The first term in the exponential of Eq. (\ref{cgf++})
consists of charge-like observables characterizing 
the state of the system for $t_{j}<t<t_{j+1}$.
This contribution can be {\it absorbed} in the diagonal 
components of L, i.e., by the substitution: 
\begin{equation}
\gamma_\alpha\rightarrow
\gamma_\alpha(\xi_1,\xi_2,\cdots)\equiv 
\gamma_\alpha+\sum_f n_f(\alpha)\xi_f
\label{subs1}
\end{equation}
If one considers the distribution of charge in our molecular quantum dot,
this reduces simply to the replacement,
\begin{equation}
\gamma=
\left(\begin{array}{cc}
\gamma_0 & 0 \\
0 & \gamma_1
\end{array}\right)
\rightarrow
\gamma(\xi)=
\left(\begin{array}{cc}
\gamma_0 & 0 \\
0 & \gamma_1+\xi
\end{array}\right).
\nonumber
\end{equation}

The second term in the exponential of Eq. (\ref{cgf++}) is, 
on the other hand, composed of
current-like observables which are associated with
transitions $\Gamma_{\alpha\beta}$ between 
different states.
In the perturbative expansion with respect to such jumps, 
these appear only at specific times $t=t_j$.
Eq. (\ref{jumps}) suggests that these contribution can be absorbed,
in the off-diagonal part $\Gamma$ as follows:
\begin{equation}
\Gamma^{(\chi)}_{\{\Delta n_f\}}\rightarrow 
\Gamma^{(\chi)}_{\{\Delta n_f\}}(\eta_1,\eta_2,\cdots)\equiv 
\Gamma^{(\chi)}_{\{\Delta n_f\}}\exp\Big[i \sum_f\Delta n_f \eta_f\Big].
\label{subs2}
\end{equation}
Note that only a part of $\Gamma$ associated with the
lead $\chi$ which is involved in the measurement, i.e.,
$\Gamma^{(\chi)}$ is subjective to the shift rule.
In our molecular quantum dot, $\Gamma$ has the following simple
structure,
\begin{equation}
\Gamma=
\left(\begin{array}{cc}
0 & \Gamma_{01}^{(L)} \\
\Gamma_{10}^{(L)} & 0
\end{array}\right)+
\left(\begin{array}{cc}
0 & \Gamma_{01}^{(R)} \\
\Gamma_{10}^{(R)} & 0
\end{array}\right).
\nonumber
\end{equation}
One also adopts the convention to measure the {\it net} current which
flows {\it out of} the left reservoir onto the quantum dot.
Then, the above replacement leads to,
\begin{equation}
\Gamma\rightarrow
\Gamma(\eta)=
\left(\begin{array}{cc}
0 & \Gamma_{01}^{(L)}e^{-i\eta}\\
\Gamma_{10}^{(L)}e^{i\eta} & 0
\end{array}\right)+
\left(\begin{array}{cc}
0 & \Gamma_{01}^{(R)} \\
\Gamma_{10}^{(R)} & 0
\end{array}\right).
\nonumber
\end{equation}

The above arguments show that calculation of the CGF is 
actually equivalent to solving a stochastic problem, 
specified by the following modified master equation,
\begin{eqnarray}
{d\over dt}p(\xi,\eta;t)&=&L(\xi,\eta)p(\xi,\eta;t),
\\
L(\xi,\eta)&=&\gamma (\xi_1,\xi_2,\cdots) +\Gamma (\eta_1,\eta_2,\cdots),
\label{masmod}
\end{eqnarray}
where $p(\xi,\eta;t)$ is a vector:
\begin{equation}
p(\xi,\eta;t)=\left(
\begin{array}{c}
p_{\alpha_1}(\xi,\eta;t) \\
p_{\alpha_2}(\xi,\eta;t) \\
\vdots 
\end{array}
\right).
\nonumber
\end{equation}
Its $(\xi,\eta)$-dependence is only written symbolically, but its
implication would be clear:
$(\xi,\eta)=(\xi_1,\xi_2,\cdots; \eta_1,\eta_2,\cdots)$.
For a given initial condition $p(\xi,\eta;t_0)=p(t_0)$,
the CGF is simply related to $p(\xi,\eta;t)$ as,
\begin{equation}
\exp\Big[\Omega(\xi,\eta)\Big]
=\sum_\alpha p_\alpha(\xi,\eta;t_0+\tau).
\nonumber
\end{equation}
The modified master equation (\ref{masmod})
can also be integrated formally to give,
\begin{equation}
p(\xi,\eta;t)=\exp\Big[(t-t_0)L(\xi,\eta)\Big]p(t_0).
\nonumber
\end{equation}
Thus for a long measurement time $\tau$, the maximal eigenvalue
$\lambda_M(\xi,\eta)$ of the modified rate matrix $L(\xi,\eta)$ 
dominates, i.e.,
\begin{equation}
\Omega(\xi,\eta)\simeq\tau\lambda_M (\xi,\eta).
\label{cgfmax}
\end{equation}
In the limit of vanishing counting fields: 
$\xi,\eta \rightarrow  0$,
one can verify that $\lambda_M (\xi,\eta) \rightarrow  0$,
the latter corresponding to the stationary state solution.

We have thus established a recipe for obtaining FCS.
We start with a master equation, Eq. (\ref{mas}),
controlled by a rate matrix with its explicit elements calculated 
through the Fermi's golden rule.
We then make the replacements (\ref{subs1},\ref{subs2}) to
define a new stochastic problem in terms of a modified master
equation with counting fields: Eq. (\ref{masmod}).
Finally, our task reduces to finding the maximal eigenvalue, 
$\lambda_M (\xi,\eta)$
of the modified rate matrix $L(\xi,\eta)$ .
The eigenspace corresponding to 
$\lambda_M (\xi,\eta)$
is smoothly connected to the stationary state 
in the limit of vanishing counting fields.


\section{The analytic solution}

The master equation describing an isotropic molecular quantum dot magnet
in the incoherent tunneling regime was derived in Sec. II.
A prescription for calculating FCS of transport through a quantum
dot starting from such a master equation was given in Sec. III.
The purpose of this section is to apply the formalism developed
in Sec. III to the specific case of our molecular quantum 
dot magnet and obtain an analytic expression for the FCS of charge and current.

In practice we have to solve an eigenvalue problem for the {\it modified 
master equation with counting fields}, 
which can be constructed by making suitable replacements of the
diagonal and off-diagonal components as Eqs.~(\ref{subs1}) and (\ref{subs2}).
In this work, we  will concentrate on the distribution of 
charge and current in our molecular quantum dot magnet.
Our modified rate matrix is, therefore, a function of only
two counting fields, say, $\xi$ and $\eta$, i.e.,
\begin{equation}
L(\xi,\eta)=
\left(\begin{array}{cc}
\gamma_0 & \Gamma_{01}^{(L)}e^{-i\eta}+\Gamma_{01}^{(R)}\\
\Gamma_{10}^{(L)}e^{i\eta}+\Gamma_{10}^{(R)} & \gamma_1+\xi
\end{array}\right)
\label{modratemat}
\end{equation}
Its eigenvalues and the corresponding eigenvectors are also
functions of $\xi$ and $\eta$:
\begin{equation}
L(\xi,\eta) u(\xi,\eta)=\lambda(\xi,\eta)u(\xi,\eta).
\label{eigen}
\end{equation}
The FCS of our problem is obtained as a cumulant generating function 
of the charge-current joint correlation functions,
which is identified as the maximal eigenvalue
$\lambda(\xi,\eta)$
of the modified rate matrix, Eq. (\ref{modratemat}).
In the limit of vanishing counting fields $\xi,\eta \rightarrow 0$,
the eigenstate $u(\xi,\eta)$ corresponding to $\lambda(\xi,\eta)$,
collapses smoothly on the stationary state $u_0$, i.e.,
\begin{eqnarray}
\lim_{\xi,\eta \rightarrow 0}u(\xi,\eta)=u(0,0)=u_0,
\nonumber \\ 
\lim_{\xi,\eta \rightarrow 0}\lambda(\xi,\eta)=\lambda(0,0)=0. 
\label{solstat}
\end{eqnarray}
In the same limit, Eq. (\ref{modratemat}) reduces to Eq. (\ref{ratemat}).

Obtaining FCS of a system thus does not require to solve completely
the modified master equation,
but simply to identify the maximal eigenvalue, corresponding to the
stationary state solution in the limit of vanishing counting fields.
It is still rather surprising that one has an access to an analytic
solution of such a problem for arbitrary $s$, because:
\begin{enumerate}
\item
The size of the matrix one has to diagonalize is ``large'',
e.g., $6\times 6$ for $s=1/2$, 
and $3(2s+1)\times 3(2s+1)$ for a general $s$.
\item
For arbitrary $s$, one has to diagonalize such a matrix of that size,
and it is generally not intuitive whether the obtained eigenvalues 
have a closed form as a function of $s$. 
\end{enumerate}
For $s=1/2$ the $6\times 6$ eigenvalue problem, Eqs. (\ref{eigen}) and (\ref{modratemat}),
can be treated analytically by reducing the size of the problem
down to $2\times 2$, using the identity,
\begin{equation}
\det 
\left(\begin{array}{cc}
A & B \\
C & D
\end{array}\right)
= \det (A-BD^{-1}C),
\label{identity}
\end{equation}
which is valid when $\det D\ne 0$.
Here the four block matrices correspond to different charge sectors.
One can apply the same identity to the case of arbitrary spin $s$,
in which the size of the problem is reduced from
$3(2s+1)\times 3(2s+1)$ down to $(2s+1)\times (2s+1)$,
because the identity (\ref{identity}) projects the whole physical space
onto the $(0,0)$-charge sector. 
Of course, one has to pay the "price" for this reduction of size:
the appearance of $BD^{-1}C$ term, in which information on the
original larger matrix is compressed.
The $BD^{-1}C$ term is, as it should be,
a square matrix of size $(2s+1)\times (2s+1)$
(though matrices $B$ and $C$ are generally not square).
Each row and column of this matrix
are characterized by a set of 
quantum numbers $(s_z,s'_z)$.
As a consequence of angular momentum selection rules, 
the matrix $BD^{-1}C$, 
or what we will call later $X_s$ or $Y_{s,j}$,
(proportional to $BD^{-1}C$),
becomes a {\it tridiagonal} matrix, i.e.,
only $(s_z,s_z)$, $(s_z,s_z+1)$ and $(s_z,s_z-1)$ elements are non-zero.

The tridiagonal matrices, $X_s$ and $Y_{s,j}$, have a very characteristic 
structure (see Appendix),
which allows us to identify one of its eigenvectors by inspection,
Eq. (\ref{xseigen}) or (\ref{yseigen}).
We will see that this eigenvector gives also a solution of
Eqs. (\ref{eigen}) and (\ref{modratemat}), which turns out, {\it accidentally}, 
to be the right solution satisfying the condition (\ref{solstat}).
The matrices $X_s$ and $Y_{s,j}$ consist of Clebsch-Gordan coefficients at
fourth order. The above characteristic feature of those matrices are
actually due to some unconventional quartic relations among
Clebsch-Gordan coefficients, which are summarized in the Appendix.

\begin{figure}[ht]
\includegraphics[width=1.0 \columnwidth]{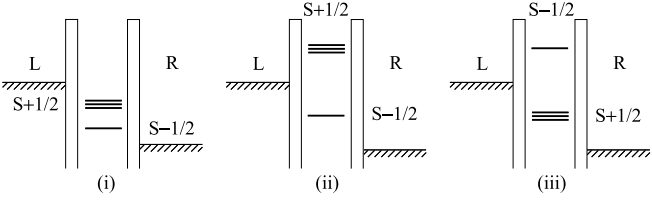}
\caption{
The dot states and the Fermi window:
(i) two spin subsectors,
(ii) only one spin subsector ($j=s-1/2$),
in the bias window.
(iii) similar to (ii), but here the "triplet"
($j=s+1/2$) subsector is in the bias window.
The figure is drawn for $s=1/2$.}
\label{fig:3states}
\end{figure}

The explicit matrix elements of Eq. (\ref{ratemat})
is given as four constituent block matrices $A-D$
in Eqs. (\ref{b})-(\ref{a}).
These formulas contain Fermi distribution functions, and therefore,
depend on the relative positions of dot levels with respect to 
the chemical potential of two leads.
At zero temperature such formulas can be further specified 
in specific situations which apply when the temperature is 
small compared to the voltage bias: (i) two spin sectors 
in the bias window,
(ii) only one spin sector in the bias window
(see Fig.~\ref{fig:3states}).

\subsection{Two spin subsectors in the bias window}

We first consider the case of two spin subsectors 
in the bias window,
$\mu_L > \left(\epsilon_{s+1/2},\epsilon_{s-1/2}\right) > \mu_R$.
At zero temperature,
Eqs. (\ref{b})-(\ref{a}) become
\begin{eqnarray}
A(s_z,s_z)&=&-\sum_{j=s\pm 1/2}\sum_{j_z=-j,\cdots,j}C_j(j_z,s_z)=-2\Gamma_L,
\nonumber \\
B_j(s_z,j_z)&=&\Gamma_R|\langle s,s_z;1/2,\sigma_z|j,j_z\rangle|^2,
\nonumber \\
C_j(j_z,s'_z)&=&\Gamma_L|\langle s,s'_z;1/2,\sigma'_z|j,j_z\rangle|^2,
\nonumber \\
D_j(j_z,j_z)&=&-\sum_{s_z=-s,\cdots,s}B_j(s_z,j_z)
=-\Gamma_R.
\label{abcd}
\end{eqnarray}
The rate matrix (\ref{ratemat}) has the simple form:
\begin{equation}
L(\xi,\eta)=
\left(\begin{array}{cc}
A & B \\
Ce^{i\eta} & D+\xi I
\end{array}\right).
\nonumber
\end{equation}
With the use of identity (\ref{identity}),
the eigenvalue problem defined by Eqs. (\ref{eigen}) and (\ref{modratemat}) 
reduces to finding $\lambda$ which satisfies,
\begin{equation}
\det \big[ A-\lambda I-B\{D+(\xi-\lambda)I\}^{-1}Ce^{i\eta} \big]=0.
\nonumber
\end{equation}
As mentioned before, the $BD^{-1}C$-like term 
is a $(2s+1)\times (2s+1)$, tridiagonal matrix with finite 
matrix elements only at $(s_z,s_z)$, $(s_z,s_z+1)$ and $(s_z,s_z-1)$.
Recalling Eq. (\ref{abcd}), one finds,
\begin{equation}
B\{D+(\xi-\lambda)I\}^{-1}Ce^{i\eta}
=-{\Gamma_L \Gamma_R e^{i\eta}\over \Gamma_R+\lambda-\xi} X_s,
\nonumber
\end{equation}
where $X_s$ is a $(2s+1)\times (2s+1)$ matrix, 
whose explicit matrix elements are,
\begin{eqnarray}
X_s (s_z,s'_z)&=&
\sum_{j=s\pm 1/2}\sum_{j_z=-j,\cdots,j}
|\langle s,s_z;1/2,\sigma_z|j,j_z\rangle|^2
\nonumber \\
&\times&
|\langle s,s'_z;1/2,\sigma'_z|j,j_z\rangle|^2,
\label{xs}
\end{eqnarray}
The reduced eigenvalue equation now reads,
\begin{equation}
\det \big[ (-2\Gamma_L-\lambda) I
+{\Gamma_L\Gamma_R e^{i\eta}\over \Gamma_R+\lambda-\xi} X_s\big]=0.
\nonumber
\end{equation}
As is explained in the Appendix, this matrix $X_s$ 
has the striking property:
\begin{equation}
X_s \left(\begin{array}{c} 1 \\ 1 \\ \vdots \\ 1 \end{array}\right) 
=2 \left(\begin{array}{c}  1 \\ 1 \\ \vdots \\ 1 \end{array}\right),
\label{xseigen}
\end{equation}
valid for arbitrary $s$.
The vector $v_0=(1,1,\cdots,1)^T$ is thus an eigenvector of $X_s$
with an eigenvalue $2$ for all $s$, 
but it is also an eigenvector of $A-\lambda I$.
This means that if $\lambda=\lambda(\xi,\eta)$
is a solution of the quadratic equation,
\begin{equation}
{(2\Gamma_L+\lambda)(\Gamma_R+\lambda-\xi)
\over \Gamma_L \Gamma_R e^{i\eta}}=2,
\label{quad}
\end{equation}
then $v_0$ satisfies,
\begin{equation}
\big[ A-\lambda I-B \{ D+(\xi-\lambda) I\}^{-1}C e^{i\eta} \big]v_0=0,
\nonumber
\end{equation}
Consequently, this solution $\lambda$ is a solution of the 
original eigenvalue problem defined by Eqs. (\ref{modratemat}) and (\ref{eigen}).
However, at this point it is still only ``a'' solution of the original 
eigenvalue problem. 
The remarkable feature of Eq. (\ref{quad}) is that
in the limit of $\xi,\eta\rightarrow 0$, if one takes also
the limit of vanishing counting fields:
$\lambda\rightarrow 0$, then on the left hand side,
the numerator and the denominator cancel each other simply 
to give a numerical factor 2, which coincides the right hand side
of the equation.
This implies that a solution of the quadratic equation
(\ref{quad}) is smoothly connected to the stationary state
solution, which satisfies Eq. (\ref{solstat}).
Indeed, one of the solutions of Eq. (\ref{quad}):
\begin{eqnarray}
\lambda(\xi,\eta)&=&-{2\Gamma_L+\Gamma_R-\xi\over 2}
\nonumber \\
&+&{1\over 2}
\sqrt{(2\Gamma_L-\Gamma_R+\xi)^2+8\Gamma_L\Gamma_R e^{i\eta}}
\label{cgf2}
\end{eqnarray}
satisfies Eq. (\ref{solstat}). This means that
Eq. (\ref{cgf2}) is actually the desired cumulant generating function
yielding the FCS.
Thus we are able to {\it identify} the suitable eigenvalue
of the problem formulated as Eqs. (\ref{eigen}) and (\ref{modratemat}) 
simply by inspection.
The obtained result (\ref{cgf2}) for general $s$ is identical to 
the $s=1/2$ case, because the corresponding eigenvalue of 
$X_s$, i.e., Eq. (\ref{xseigen}) is not dependent on $s$.
Note that Eq. (\ref{cgf2}) is also equivalent to the
the CGF calculated for spinful conduction electrons in the absence
of local (molecular) spin.
\cite{bagrets}

\subsection{Only one spin subsector in the bias window}

Consider now the case of only one spin subsector, say,
$j=s-1/2$ in the bias window, i.e., 
$\epsilon_{s+1/2} > \mu_L > \epsilon_{s-1/2} > \mu_R$.
This corresponds to the case (ii) in Fig.~\ref{fig:3states}.
In this case, Eqs. (\ref{b})-(\ref{a})
reduce, at zero temperature, to: 
\begin{eqnarray}
A(s_z,s_z)&=&-{2s\over 2s+1}\Gamma_L,
\nonumber \\
B_{s-1/2}(s_z,j_z)&=&\Gamma_R|\langle s,s_z;1/2,\sigma_z|j=s-1/2,j_z\rangle|^2,
\nonumber \\
B_{s+1/2}(s_z,j_z)&=&(\Gamma_L+\Gamma_R)
\nonumber \\
&\times&
|\langle s,s_z;1/2,\sigma_z|j=s+1/2,j_z\rangle|^2,
\label{b1}
\nonumber \\
C_{s-1/2}(j_z,s'_z)&=&\Gamma_L|\langle s,s'_z;1/2,\sigma'_z|j=s-1/2,j_z\rangle|^2,
\nonumber \\ 
C_{s+1/2}(j_z,s'_z)&=&0,
\label{c1}
\nonumber \\
D_{s-1/2}(j_z,j_z)&=&-\Gamma_R,
\nonumber \\
D_{s+1/2}(j_z,j_z)&=&-\Gamma_L-\Gamma_R,
\nonumber
\end{eqnarray}
where $\sigma_z$ and $\sigma'_z$ are not independent variables, but
they are determined automatically by the constraints: 
$j_z=s_z+\sigma_z$ and $j_z=s'_z+\sigma'_z$, respectively.
The modified rate matrix can be obtained as Eq. (\ref{modratemat}).
Again we use the identity, Eq. (\ref{identity}).
Note that $C_{s+1/2}=0$. This allows us to rewrite the
$BD^{-1}C$ term as,
\begin{equation}
-{B_{s-1/2}C_{s-1/2}e^{i\eta}\over \Gamma_R+\lambda-\xi}
=-{\Gamma_L \Gamma_R e^{i\eta}\over \Gamma_R+\lambda-\xi} Y_{s,s-1/2},
\nonumber
\end{equation}
where
$Y_{s,j}$ is a $(2s+1)\times (2s+1)$ square matrix, 
whose $(s_z,s'_z)$-elements are,
\begin{eqnarray}
Y_{s,j} (s_z,s'_z)=
\sum_{j_z=-j,\cdots,j}
|\langle s,s_z;1/2,\sigma_z|j,j_z\rangle|^2
\nonumber \\
\times
|\langle s,s'_z;1/2,\sigma'_z|j,j_z\rangle|^2.
\label{ys}
\end{eqnarray}
The eigenvalue equation reads,
\begin{equation}
\det \Bigg[ \left( -{2s\over 2s+1}\Gamma_L-\lambda\right) I
+{\Gamma_L\Gamma_R e^{i\eta}\over \Gamma_R+\lambda-\xi} Y_{s,s-1/2}\Bigg]=0.
\label{redeigen1}
\end{equation}

Another situation to be considered in parallel is the case
of only $j=s+1/2$ spin sector in the bias window, 
i.e., $\epsilon_{s-1/2} > \mu_L > \epsilon_{s+1/2} > \mu_R$.
This corresponds to the case (iii) in Fig.~\ref{fig:3states}.
In this case, Eqs. (\ref{b})-(\ref{a}) reduce,
at zero temperature, to,
\begin{eqnarray}
A(s_z,s_z)&=&-{2s+2\over 2s+1}\Gamma_L,
\nonumber \\
B_{s-1/2}(s_z,j_z)&=&(\Gamma_L+\Gamma_R)
\nonumber \\
&\times&
|\langle s,s_z;1/2,\sigma_z|j=s-1/2,j_z\rangle|^2,
\nonumber \\
B_{s+1/2}(s_z,j_z)&=&\Gamma_R|\langle s,s_z;1/2,\sigma_z|j=s+1/2,j_z\rangle|^2,
\nonumber \\
C_{s-1/2}(j_z,s'_z)&=&0,
\nonumber \\ 
C_{s+1/2}(j_z,s'_z)&=&\Gamma_L|\langle s,s'_z;1/2,\sigma'_z|j=s+1/2,j_z\rangle|^2,
\label{c1''}
\nonumber \\
D_{s-1/2}(j_z,j_z)&=&-\Gamma_L-\Gamma_R,
\nonumber \\
D_{s+1/2}(j_z,j_z)&=&-\Gamma_R.
\nonumber
\end{eqnarray}
The modified rate matrix can be constructed accordingly.
Note that this time $C_{s-1/2}=0$. After the use of identity (\ref{identity}), 
the reduced eigenvalue equation reads,
\begin{equation}
\det \Bigg[ \left( -{2s+2\over 2s+1}\Gamma_L-\lambda\right) I
+{\Gamma_L\Gamma_R e^{i\eta}\over \Gamma_R+\lambda-\xi} Y_{s,s+1/2}\Bigg]=0.
\label{redeigen2}
\end{equation}

We now attempt to identify the maximal eigenvalue $\lambda(\xi,\eta)$,
following the same logic as the case of two spin subsectors in the
bias window.
As expected, the matrix $Y_{s,j}$ has the following characterizing
feature:
\begin{equation}
Y_{s,j} \left(\begin{array}{c} 1 \\ 1 \\ \vdots \\ 1 \end{array}\right) 
={2j+1\over 2s+1}\left(\begin{array}{c}  1 \\ 1 \\ \vdots \\ 1 \end{array}\right).
\label{yseigen}
\end{equation}
This is explained in the Appendix.
Eqs. (\ref{redeigen1}),(\ref{redeigen2}) and (\ref{yseigen}) 
suggests that $\lambda(\xi,\eta)$ should satisfy
a quadratic equation analogous to Eq. (\ref{quad}),
\begin{equation}
{\left({2j+1\over 2s+1}\Gamma_L+\lambda\right)(\Gamma_R+\lambda-\xi)
\over \Gamma_L \Gamma_R e^{i\eta}}={2j+1\over 2s+1},
\label{quad1}
\end{equation}
then $\lambda(\xi,\eta)$ satisfies also 
Eqs. (\ref{redeigen1}) and (\ref{redeigen2}).
Of course, at this point it is still only ``a'' solution of the original 
eigenvalue problem. 
However, Eq. (\ref{quad}) has again the following remarkable property:
in the limit of vanishing counting fields,
$\xi,\eta\rightarrow 0$, if one takes also
the limit $\lambda\rightarrow 0$, then on the left hand side
the numerator cancels with the denominator to give simply
a factor ${2j+1\over 2s+1}$, 
which is {\it insensitive} to the $\Gamma$'s, and
coincides with the right hand side of the equation.
This a rather unexpected coincidence
because the factor ${2j+1\over 2s+1}$ on the left hand side of the equation
comes from a quadratic relation (\ref{szquad}),
whereas the same ${2j+1\over 2s+1}$ factor on the right hand side
originates from a quartic relation (\ref{quart}).
Indeed, one of the solutions of Eq. (\ref{quad1}),
\begin{eqnarray}
&&\lambda(\xi,\eta)=-{{2j+1\over 2s+1}\Gamma_L+\Gamma_R-\xi\over 2}
\nonumber \\
&&+{1\over 2}\sqrt{\left({2j+1\over 2s+1}\Gamma_L-\Gamma_R+\xi\right)^2+
4{2j+1\over 2s+1}\Gamma_L\Gamma_R e^{i\eta}},
\nonumber \\
\label{cgf1}
\end{eqnarray}
satisfies Eq. (\ref{solstat}).
We are thus able to solve the problem simultaneously for the
two cases, i.e., either of the spin subsectors ($j=s\mp 1/2$) 
in the bias window, 
corresponding to the cases (ii) and (iii) in Fig.~\ref{fig:3states}.
Contrary to the previous case, i.e., Eq. (\ref{cgf2})
for two spin subsectors in the bias window,
result which we obtain (\ref{cgf1}) depends on $s$.
This is because the eigenvalue of Eq. (\ref{yseigen})
is dependent on both $s$ and $j$.
The generating function $\lambda(\xi,\eta)$ in Eq. (\ref{cgf1})
is also asymmetric with respect to the bare amplitudes, 
$\Gamma_L$ and $\Gamma_R$, whereas this asymmetry tends to disappear
for large spin $s$. These characteristic features of the
charge-current joint distribution function are further
discussed in the next section.

\subsection{Remarks}
We have thus successfully calculated the FCS 
for transport through a molecular quantum dot magnet 
in the incoherent tunneling regime.
The results are obtained,
in Eqs.~(\ref{cgf2}) and (\ref{cgf1}), together with Eq. (\ref{cgf})
in the form of a cumulant generating function (CGF) of charge-current joint 
correlation functions.
Here, we apply the obtained CGF for deducing lowest-order current 
correlation functions,
in order to check consistency with known results.

In the case of two spin subsectors in the bias window,
i.e., case (i) in Fig.~\ref{fig:3states},
the obtained CGF, i.e., Eq. (\ref{cgf2}) has no $s$-dependence.
The lowest order cumulants are obtained by taking the derivatives
of CGF with respect to counting fields, 
and have no $s$-dependence.
Thus the Fano factor is given for an arbitrary $s$ by,
\begin{equation}
{C_{02}\over C_{01}}={4\Gamma^2_L+\Gamma^2_R\over (2\Gamma_L+\Gamma_R)^2},
\label{c2}
\end{equation}
which coincides with the result for spinful conduction electrons and
no local spin.
\cite{struben}
The skewness normalized by $C_{01}$ reads,
\begin{equation}
{C_{03}\over C_{01}}=
\frac{16\Gamma_L^4-16\Gamma_L^3 \Gamma_R+24\Gamma_L^2 \Gamma_R^2
-4\Gamma_L \Gamma_R^3+\Gamma_R^4}
{(2\Gamma_L+\Gamma_R)^4}.
\label{c3}
\end{equation}
Note that Eqs. (\ref{c2}) and (\ref{c3}) can be rewritten
in terms of simple polynomials of a single parameter,
$x=(2\Gamma_L-\Gamma_R)^2/(4\Gamma_L\Gamma_R)$ as,
\[
{C_{02}\over C_{01}}={x+1\over x+2},\ \ \
{C_{03}\over C_{01}}={x^2+x+1\over (x+2)^2}.
\]
One might expect that the (extended) Fano factors,
$C_{04}/C_{01}, C_{05}/C_{01}, \cdots$
could be obtained 
by developing this apparently systematic series.
However, this naive expectation turns out to be an artifact at fourth order.

In the case of $j=s\pm1/2$ spin subsector in the bias window,
one has to use Eq. (\ref{cgf2}), instead of Eq. (\ref{cgf1}) as the CGF.
Notice that (\ref{cgf1}) is obtained by making the following replacement in
Eq. (\ref{cgf2}): 
$2\Gamma_L\rightarrow  {2j+1\over 2s+1}\Gamma_L$,
e.g., the Fano factor becomes in this case,
\begin{equation}
{C_{02}\over C_{01}}=
{\left({2j+1\over 2s+1}\right)^2\Gamma^2_L+\Gamma^2_R
\over \left({2j+1\over 2s+1}\Gamma_L+\Gamma_R\right)^2}.
\nonumber
\end{equation}
In the limit of large local spin $s\rightarrow\infty$, this
reproduces the result for spinless conduction electrons,
\begin{equation}
{C_{02}\over C_{01}}\rightarrow
{\Gamma^2_L+\Gamma^2_R
\over (\Gamma_L+\Gamma_R)^2}.
\end{equation}
The third order Fano factor
(skewness normalized by $C_{01}$)
is also obtained by applying the same replacement to
Eq. (\ref{c3}).

\section{Discussions}

The exact analytic expressions for the CGF, Eqs.~(\ref{cgf2}) and (\ref{cgf1}), 
contain the full information of the non-equilibrium statistical properties of the current and of the charge. 
Though the CGF and the joint probability distribution contain the same information 
as we can see from their definitions (Sec. III), 
the former is rather a mathematical tool, whereas the latter provides us with
a physical picture. 
We therefore apply an inverse Fourier transformation
to the obtained CGF, transforming it back to a joint probability distribution: 
\begin{eqnarray}
P(N,I)
=
\frac{\tau}{(2 \pi)^2} \!
\int_{-\pi}^{\pi} 
\!\!\!\!\!\!
d \eta
\int_{-\infty}^{\infty} 
\!\!\!\!\!\!
d \, \xi 
\;
{\rm e}^
{
\Omega(\xi,\eta)
- i \, \tau N \xi - i \, \tau I \eta
}
\nonumber.
\end{eqnarray}
%
The asymptotic form for $\tau \! \rightarrow \! \infty$ is obtained 
within the saddle point approximation, since $\Omega$ is proportional to $\tau$,
and thus the exponent is also proportional to $\tau$: 
$
\ln P(N,I)
\approx
\Omega(\xi^*,\eta^*)
-iN \tau \xi^*-iI \tau \eta^*,
$
where $\xi^*$ and $\eta^*$ are determined by the saddle point equations,
$N \, \tau \!=\! -i \, \partial_\xi  \, \Omega(\xi^*,\eta^*)$
and 
$I \, \tau \!=\!  -i \, \partial_\eta \, \Omega(\xi^*,\eta^*)$~\cite{bagrets}.
Combining these equations, we can produce a contour plot of $\ln P(N,I)$ 
as a parametric plot in terms of $\eta$ and $\xi$. 
The probability distribution for only one of the observables, i.e., either 
charge or current, is obtained further by fixing $\eta^*\!=\!0$ or $\xi^*\!=\!0$,
respectively.

In this section, we try to visualize the joint probability distribution, and 
discuss how a molecular spin influences statistical properties of the SET molecular
quantum dot. 
The joint distribution reveals nontrivial correlations: 
correlations among multiple current components in a multi-terminal chaotic cavity,
\cite{bagrets}
or correlations between two different observables, such as current and charge, 
\cite{utsumi_condmat}
have been investigated.
Here, we first consider a dot with symmetric tunnel junctions 
$\Gamma_L \!=\! \Gamma_R$, 
and analyze the joint distribution of charge and current under 
high and small bias voltages.
We then consider the case of a large asymmetry in the tunnel coupling,
e.g., $\Gamma_L \ll \Gamma_R$.
We will argue that much information on the molecular spin and the exchange coupling 
can be extracted from transport measurements. 
Finally, we will plot the lowest three cumulants, which are indeed being measured
in the state-of-art experiments. 
\begin{figure}[ht]
\includegraphics[width=1.0 \columnwidth]{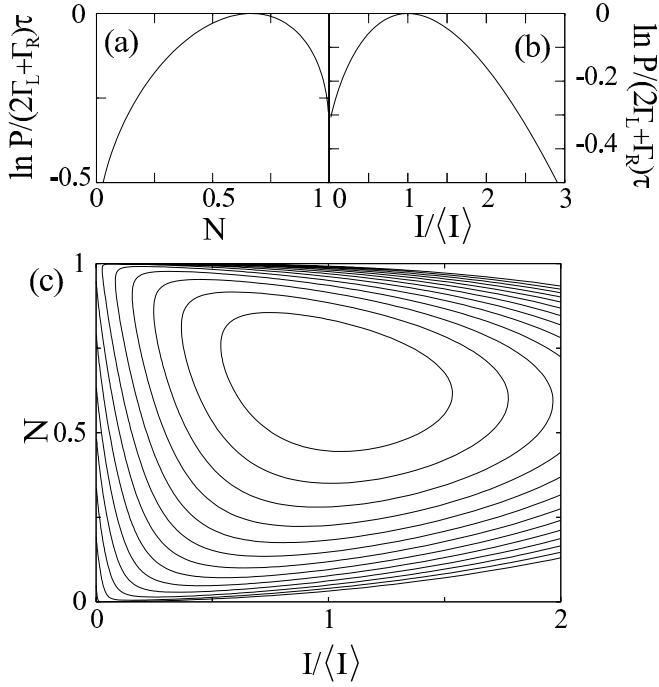}
\caption{
The probability distributions of charge (a) and current (b) for the symmetric coupling 
$\Gamma_L \!=\! \Gamma_R$. 
The bias voltage is large enough and both of the two spin subsectors are in the bias window 
[Fig.~\ref{fig:3states}~(i)].
(c) The contour plot of the non-equilibrium charge-current joint distribution $\ln P(N,I)$. 
The horizontal (vertical) axis corresponds to current $I$ (charge $N$).
The contour interval is $(2 \Gamma_L \!+\! \Gamma_R) \tau/20$. 
}
\label{fig:highbias}
\end{figure}

\subsection{Effects of local spin on the statistical properties 
for symmetric tunnel coupling $\Gamma_L \! = \! \Gamma_R$}
\label{sec1}
%
%
Let us consider a molecular quantum dot with symmetric tunnel junctions $\Gamma_L=\Gamma_R$. 
We first consider the case of a large bias voltage, in which the two spin subsectors 
are both in the bias window, 
i.e. the case (i) in Fig.~\ref{fig:3states}. 
In this case the CGF has been calculated analytically in Eq. (\ref{cgf2}).
Here, we attempt to uncover basic physical consequences, hidden in this result.
In Figs. 2(a) and 2(b),
the probability distribution functions of, respectively, charge and current
are plotted. They have been evaluated by applying the saddle-point approximation
explained above to the solution, Eq. (\ref{cgf2}).
In the two cases, the distribution function is not symmetric in regard to the
peak position, indicating a clear deviation from the Gaussian distribution.
Such asymmetric distributions were somewhat expected, 
since, e.g., in the case of charge distribution,
the average value $\langle N \rangle$ is not located at 1/2 
but at 2/3. 

In Fig. 2(c) the two distributions are shown simultaneously
as a joint distribution in the form of a contour plot. 
The distribution possesses a single peak at $I \!=\! \langle I \rangle$ and 
$N \!=\! \langle N \rangle$. 
For a given value of current $I$,
the peak position of $N$ converges to $N \!=\! \langle N \rangle  \!=\! 2/3$
as the current becomes larger ($I \! \gg \! \langle I \rangle$).
Note that the probability distributions depend neither on the size $s$ of 
the molecular spin nor the exchange coupling $g$. 
The asymmetry in the charge distribution comes from the coefficient 2 
in front of $\Gamma_L$ [Eq.~(\ref{cgf2})], 
reflecting the number of degrees of freedom of a transmitted spin. 
The truth is that Eq.~(\ref{cgf2}) is not different from the CGF in the absence of
a molecular spin, i.e., the coefficient 2 appears also in the latter case. \cite{bagrets} 
We thus conclude that at a high enough bias voltage, 
all the statistical properties will reproduce the results in the absence of a molecular spin. 

%
%

\begin{figure}[ht]
\includegraphics[width=0.75 \columnwidth]{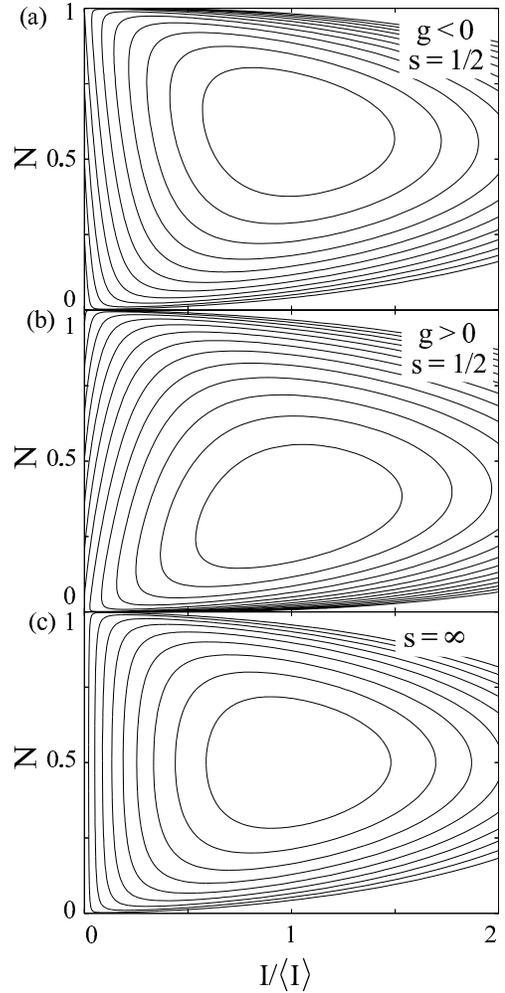}
\caption{
Contour plot of the joint probability distribution $\ln P(N,I)$ 
in the case of only one spin subsector in the bias window 
[Figs.~\ref{fig:3states} (ii) and (iii)].
The coupling to reservoirs is assumed to be symmetric, $\Gamma_L \!=\! \Gamma_R$. 
Panels (a) and (b) correspond to a small local spin $s \!=\! 1/2$,
whereas panel (c) corresponds to a large spin $s \! \rightarrow \! \infty$. 
The exchange coupling between the local and an itinerant spin
is ferromagnetic ($g\!<\!0$) in panel (a), whereas antiferromagnetic ($g\!>\!0$) 
in panel (b).  
The contour interval is 
$(\frac{2j\! + \!1}{2s\!+\!1} \Gamma_L \!+\! \Gamma_R)/20$. 
}
\label{fig:case2}
\end{figure}

We then turn our attention to the small bias voltage regime, where only one spin subsector 
is in the bias window, the cases (ii) or (iii) in Fig.~\ref{fig:3states}. 
Contour plots of the joint probability distribution obtained by using Eq.~(\ref{cgf1}) 
are plotted in Fig.~\ref{fig:case2}. 
Figs. 3(a) and  3(b) correspond to a small molecular spin $s \!=\! 1/2$.
In both ferromagnetic ($g\!<\!0$) [panel (a)], 
and antiferromagnetic ($g\!>\!0$) [panel (b)] cases, 
the charge distribution is not symmetric, i.e., typically the peak position is
away from $N \!=\! 1/2$.
The nature of this asymmetry is similar to the previous case, i.e., 
the case of large bias voltage, Fig.~\ref{fig:highbias} (c).
The only difference is that here the peak position is located at 
$\langle N \rangle \!=\! \frac{2j\! + \!1}{2s\!+\!1} \Gamma_L 
/(\frac{2j\! + \!1}{2s\!+\!1} \Gamma_L \!+\! \Gamma_R)$, 
which is 3/5 for panel (a) and 1/3 for panel (b).

On the other hand,
the distribution becomes symmetric w.r.t. a horizontal line $N=1/2$
in the limit of a large spin $s$
[panel (c)]. 
It might be a little surprise that this limit reproduces
the joint probability distribution of a spinless fermion~\cite{utsumi_condmat}. 
In physical terms this can be interpreted as follows:
{\it a large molecular spin acts as a spin bath for the electron spin transmitted through 
the quantum dot}. 
The $n=0$ charge sector is $(2s+1)$-fold degenerate, whereas
the degeneracy of $n=1$ charge sector is either $2s$ or $2s+2$.
For a large molecular spin $s$, the conduction electron spin does not feel 
the difference between the two different total angular momentum subsectors. 
The above result is counterintuitive in the sense that
naively we expect that a large molecular spin may act as a large effective Zeeman field 
for transmitted spins. 

\subsection{Asymmetry in the tunnel coupling $\Gamma_L \! \neq \! \Gamma_R$}
\label{sec2}

Here we discuss how an asymmetry in the tunnel coupling may change the statistical properties
of a SET molecular quantum dot. 
In reality, the tunneling coupling to reservoirs is unlikely to be symmetric,
because one cannot control the coupling strength of ligands to metallic leads~\cite{heersche}. 
In the extremely asymmetric limit $\Gamma_R \! \gg \! \Gamma_L$, 
the CGF of current reduces to a Poissonian form: 
\begin{eqnarray}
\Omega
\approx
\tau \, z \, \Gamma_L \, ({\rm e}^{i \eta}-1). 
\label{cgfpoisson}
\end{eqnarray}
The factor $z$ is a positive constant, which is smaller than 2: 
$z \!=\! 2$ for the case (i) and $z \!=\! (2j\!+\!1)/(2s\!+\!1)$
with $j=s\mp 1$, respectively,
for the cases (ii) and (iii) in Fig.~\ref{fig:3states}. 
When we apply a negative bias voltage, the CGFs are obtained from
Eqs.~(\ref{cgf2}) and (\ref{cgf1}) by replacing 
$\Gamma_L \! \leftrightarrow \! \Gamma_R$ 
and 
$\eta \! \rightarrow \! -\eta$ as, 
$\bar{\Omega} \! \approx \! \tau \, \Gamma_L \, ({\rm e}^{-i \eta}-1)$. 
The absolute value of the $n$-th cumulant is 
$|C_{0n}| \! \approx \! \tau z \Gamma_L$ 
for a positive bias voltage, whereas it becomes
$|\bar{C}_{0n}| \! \approx \!  \tau \Gamma_L$ 
for a negative bias voltage. 
Therefore, from the ratio $|C_{0n}/\bar{C}_{0n}|$,
we can estimate the factor $z$, and consequently, 
obtain some information on the molecular spin and the exchange coupling. 
Such a method was previously proposed \cite{Akera} for a multi-orbital quantum dot.

\begin{figure}[h]
\includegraphics[width=0.75 \columnwidth]{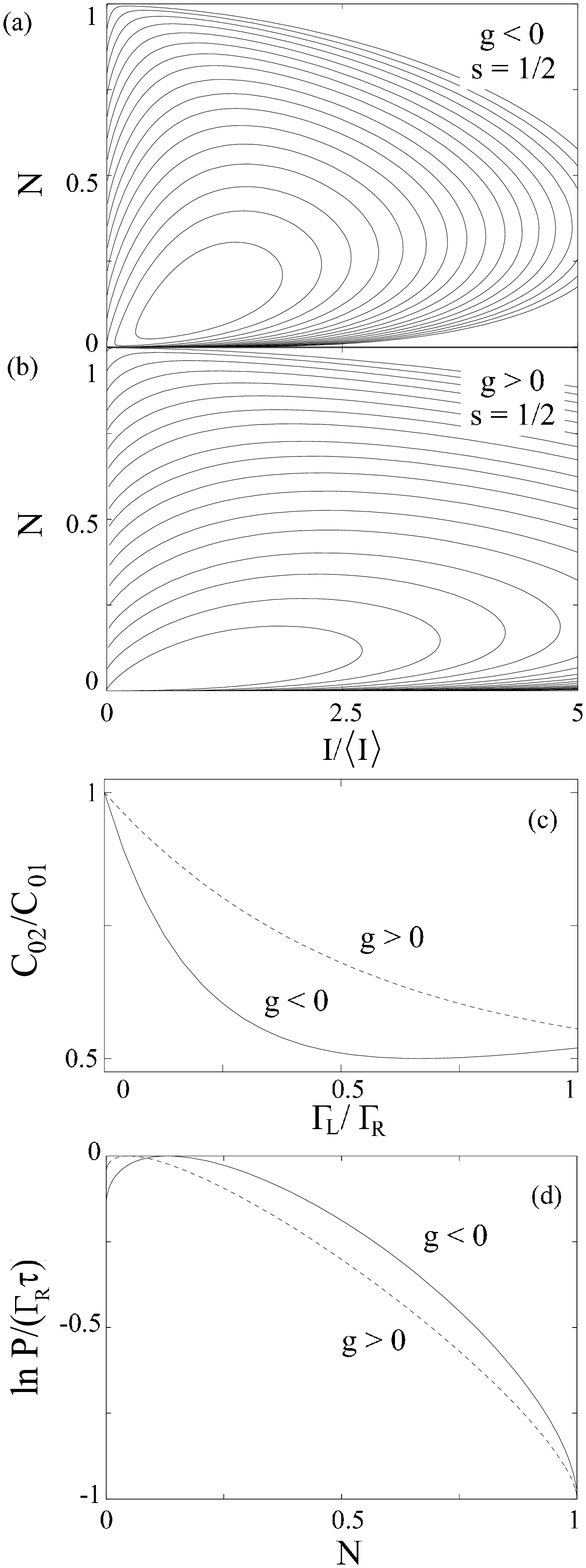}
\caption{
Contour plots of $\ln P$ for a small local spin $s=1/2$ for the asymmetric coupling 
$\Gamma_R \!=\! 10 \Gamma_L$. 
The ferromagnetic ($g\!<\!0$) and antiferromagnetic ($g\!>\!0$) coupling cases are 
shown in panels (a) and (b), respectively. 
The contour interval is $(\Gamma_L \!+\! \Gamma_R)/20$. 
(c) Fano factor as a function of the asymmetry factor $\Gamma_L/\Gamma_R$ and 
(d) Probability distributions of charge for antiferromagnetic and ferromagnetic couplings. 
}
\label{fig:asymmetry}
\end{figure}

Figs. 4(a) and 4(b) are the joint probability distribution for 
$\Gamma_L/\Gamma_R \!=\! 0.1$. 
A strong asymmetry around 
$N \!=\! \langle N \rangle \!=\! z \Gamma_L/(z \Gamma_L \!+\! \Gamma_R)$ 
is observed both for the antiferromagnetic (a) and the ferromagnetic (b) cases. 
A longer tail in the panel (b) implies that the correlations among tunneling processes 
are weaker for antiferromagnetic coupling. 
The measure of such correlations, the Fano factor $C_{02}/C_{01}$, 
as a function of the asymmetry ratio $\Gamma_L/\Gamma_R$ is shown in 
Fig.~\ref{fig:asymmetry} (c). 
We can observe that even for a small asymmetry ratio $\Gamma_L/\Gamma_R \!=\! 0.1$, 
the Fano factor is still smaller than 1, 
and a fair amount of difference remains between the two cases. 

%
%

For the charge distribution, around $I \! \approx \! 0$, we observe equidistant 
contours in panels (a) and (b), which implies that the charge is exponentially distributed. 
Actually the charge distribution for both cases roughly follows 
$\ln P(N) \! \propto \! - \tau \Gamma_R N$ [Figure~\ref{fig:asymmetry} (d)]. 
A similar exponential distribution appears as an equilibrium distribution of charge 
when the dot level is far away from either of the two lead chemical potentials~\cite{utsumi_condmat}. 
It is shown that the exponent depends simply on the decay rate of the excited state. 
In the present case, since the stationary state is fairly approximated by the empty state, 
this reduces to the outgoing tunneling rate $\Gamma_R$ of an electron in the occupied state
to the right reservoir.

%
%

\subsection{Relation to experiments}

Up to now, we have shown contour plots of the probability distribution of charge and
current. 
For a semiconducting quantum dot, it has become possible to measure the statistical probability 
distribution of current in the incoherent tunneling regime~\cite{fujisawa,gustavsson_ensslin}. 
Such measurements are based on the real-time counting of single electron jumps using an on-chip quantum point 
contact charge detectors. At the present stage, it may
be rather challenging to apply the same technique for a single-molecular 
device with the same precision, but it might be still possible to measure the lowest three cumulants, 
since the skewness for a tunnel junction can be measured by the state-of-art experiments.
\cite{reulet,reznikov}

We have shown explicitly the lowest order cumulants in Sec. IV C.
For a large bias voltage, they are given, e.g., in Eqs. (\ref{c2}) and (\ref{c3}).
For a small bias voltage, the (generalized) Fano factors were
obtained by making the replacement,
$2\Gamma_L\rightarrow  z\Gamma_L={2j+1\over 2s+1}\Gamma_L$,
in Eqs. (\ref{c2}) and (\ref{c3}).
Fig. 5 shows the bias voltage dependence of the lowest three cumulants
for a large molecular spin $s \!=\! 10$, 
which is approximately the size of the magnetic moment of ${\rm Mn_{12}}$.
\cite{heersche}
In this case, as the factor $z={2j+1\over 2s+1}=20/21$ (for $j=s-1/2$) is close to 1, 
the molecular spin is expected to act as a {\it spin bath},
reproducing characteristic features of the transport of spinless fermions. 
It is assumed that the tunnel coupling is asymmetric, $\Gamma_R \!=\! 5 \Gamma_L$.
We also focus on the antiferromagnetic ($g>0$) case, in which
the "triplet" level ($j=s+1/2$),
$\epsilon_{s+1/2}=\epsilon_{\rm dot}+ g s/2$ is above the "singlet" level ($j=s-1/2$),
$\epsilon_{s-1/2}=\epsilon_{\rm dot}- g (s+1)/2$.
In equilibrium, the singlet subsector is tuned to be at the lead chemical potential
level,  $\mu_L=\epsilon_{s-1/2}=\mu_R=0$. 
We then apply either a positive [panels (a-1) and (b-1)]
or a negative [panel (a-2) and (b-2)] bias voltage.
The singlet level always stays in the bias window, i.e.,
either $\mu_L>\epsilon_{s-1/2}>\mu_R$ [panels (a-1) and (b-1)]
or $\mu_L<\epsilon_{s-1/2}<\mu_R$ [panels (a-2) and (b-2)]
is realized.
Note that in Fig. 5 the origin of energy is taken, for simplicity, to be at the singlet
level, i.e., $\epsilon_{s-1/2}=0$ is always satisfied.

The expected cumulant-voltage characteristics for a positive and for 
a negative bias voltage are depicted in panels (a-1) and (a-2) of Fig. 5. 
At a small bias voltage, 
$|\mu_{L,R}| \! < \! \epsilon_{s+1/2} \!-\! \epsilon_{s-1/2} \! =\! g (s \!+\! 1/2)$, 
the absolute value of various cumulants are insensitive to the sign of bias voltage 
[This case realizes the situation of Fig.~\ref{fig:highbias} (c)]. 
On the other hand, when the bias voltage is large, 
$|\mu_{L,R}| \! > \! g (s \!+\! 1/2)$, 
each cumulant takes different values depending on whether the bias is 
either positive $|\mu_L| \! > \! g (s \!+\! 1/2)$, or negative 
$|\mu_R| \! > \! g (s \!+\! 1/2)$ 
[This case realizes the situation of Fig.~\ref{fig:case2} (c)]. 
Such a feature could be a smoking gun of the transport of spinless fermions. 
Panels (b-1) and (b-2) show Fano factors corresponding, respectively, 
to the panels (a-1) and (a-2). 
At a large bias voltage $|\mu_{L,R}| \! > \! g (s \!+\! 1/2)$, 
the suppression of Fano factors is stronger for a positive [panel (b-1)]
than for a negative [panel (b-2)] bias voltage. 
This can be understood as follows.
If one compares the large bias regime of panels (a-1) and (a-2),
one notices that the absolute value of current, i.e., $C_{01}$
is smaller in this regime in panel (a-2). 
This implies that tunneling events are less correlated
when the bias is large and negative, leading to smaller Fano factors
in panel (b-2).
This is rather a trivial fact, but subjective to a direct experimental check. 

\begin{figure}[ht]
\includegraphics[width=1.0 \columnwidth]{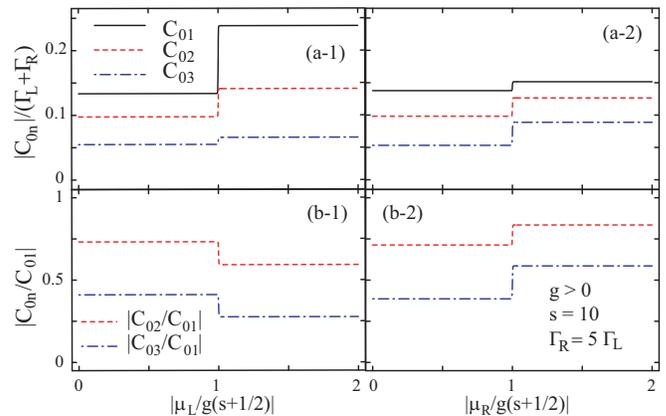}
\caption{
The absolute value of the lowest three cumulants, 
$C_{01}$, $C_{02}$ and $C_{03}$,
for a molecular spin $s \!=\! 10$ 
and an asymmetric tunnel coupling ($\Gamma_R \!=\! 5 \Gamma_L$). 
We also assumed that the exchange interaction is antiferromagnetic ($g\!>\!0$).
The origin of energy is taken to be at the "singlet" level, $\epsilon_{s-1/2}=0$.
Panels (a-1) and (a-2) corresponds, respectively, to a positive and 
to a negative bias voltage. 
In panels (b-1) and (b-2), the corresponding Fano factors 
$C_{02}/C_{01}$ and $C_{03}/C_{01}$ are plotted. 
}
\label{fig:experiment}
\end{figure}

Finally, it should be noted again that the magnetic anisotropy of the molecule was neglected
throughout the present paper. 
The extension of our theory to account for a large anisotropy comparable to 
the exchange coupling is beyond our scope here. 


\section{Conclusions}

We have thus studied the full counting statistics (FCS) for a molecular
quantum dot magnet, a quantum dot with an intrinsic molecular spin ($s$)
degrees of freedom, which is coupled to the conduction electron
spin ($\sigma$) via an exchange interaction ($s\cdot\sigma$).
Such a molecular quantum dot magnet is now experimentally available,
and carries generally a large molecular spin $s\sim 10$.
\cite{heersche}
We applied the master equation approach to FCS 
\cite{bagrets}
to our molecular quantum dot problem, 
by considering the incoherent tunneling regime.
In this approach the cumulant generating function (CGF) for FCS
is obtained by solving an eigenvalue problem associated with
a modified master equation with counting fields.
We also assumed that the molecular quantum dot was in the strong
Coulomb blockade regime ($U\rightarrow\infty$) so that the
dot state has only two charge sectors $n=0,1$.
A standard algebraic identity, Eq. (\ref{identity}), allowed us
to reduce the size of the eigenvalue problem
from $3(2s+1)\times3(2s+1)$ down to $(2s+1)\times (2s+1)$
by projecting the relevant physical space onto the charge $(0,0)$-sector.
This was, of course, simply a rewriting of the original problem.
In the new representation, Clebsch-Gordan coefficients appear at
fourth order.
We established a few identities among such fourth order coefficients, 
which allowed us to identify one solution of the eigenvalue problem,
which turned out to be the correct eigenvalue,
representing the CGF for FCS.
Thus we obtained an analytic expression for the CGF 
{\it as a function of the local spin $s$}
for different configurations of the dot states and of the two leads
in regard to their relative energy levels (Fig.~\ref{fig:3states}). 

Based on the obtained analytic expressions, we also developed
numerical analysis, e.g.,
contour plots of the charge-current joint distribution function.
For a small local spin, e.g., $s=1/2$,
the obtained contour plots show clearly an asymmetry in the distribution
of the charge, i.e., between $n=0$ and $n=1$. 
We then demonstrated that this asymmetry grows differently
for different cases in Fig.~\ref{fig:3states} mentioned above.
In particular, this asymmetry tends to disappear as the local spin $s$ 
becomes large, in cases (ii) and (iii) of Fig.~\ref{fig:3states}, 
indicating that the local spin plays the role of a {\it spin bath} for the
conduction electron.
Of course, as emphasized in Sec. 5, it is rather counterintuitive to recover
such characteristic transport properties of a {\it spinless fermion}
in this large $s$ limit.
We further demonstrated that
such characteristic features of the transport of a spinless
fermion, which we believe to have a chance to be realized in reality, 
e.g., in a Mn$_{12}$ molecular quantum dot,
are subjective to a direct experimental check through the
bias voltage dependence of lowest-order cumulants.

It is interesting to extend our analysis to the case of leads of 
different nature,
such as ferromagnetic leads or superconducting leads.
The range of validity of this work is limited to the incoherent
tunneling regime, whereas one might naturally wonder
what happens when the coupling between the dot and the leads become
stronger.
Kondo type physics in such a regime is particularly interesting,
\cite{romeike_kondo}
because of the coexistence of an intrinsic magnetic impurity
(molecular spin) and the quantum dot, 
the latter known to show enhancement of conductance due to
Kondo mechanism at low temperatures.
Certainly one has to go beyond the master equation approach to
FCS in order to treat FCS in such a regime.

\acknowledgements
We are grateful to JSPS Invitation Fellowship for Research in Japan. 
K.I. and Y.U. are supported by RIKEN as Special Postdoctoral Researcher.
T.M. also acknowledges support from French CNRS A.C. Nanosciences grant
NR0114. He thanks the members of Condensed Matter Theory Laboratory at 
RIKEN for their hospitality. 

\vspace{1cm}
\begin{center}
{\bf\large Appendices}
\end{center}
\begin{appendix}

\section{Useful relations among Clebsch-Gordan coefficients}

For the calculation of FCS for a model with an arbitrary spin $s$
we have used extensively the Clebsch-Gordan algebra.
We, therefore, summarize in the following useful relations among 
Clebsch-Gordan coefficients.
We start with conventional quadratic relations which are sometimes
discussed in standard quantum mechanics textbooks,
and then proceed to unconventional quartic relations which we
discovered as a byproduct of this research.

\subsection{Quadratic relations}

Let us summarize the quadratic identities on the Clebsch-Gordan 
coefficients,
$\langle s,s_z;1/2,\sigma_z|j,j_z\rangle$,
for a given $s$ and $\sigma=1/2$.
Angular momentum selection rule allows only two values of
$j=s\pm 1/2$.
Because of the constraint $j_z=s_z+\sigma_z$,
{\it only two of the three parameters}, $s_z$, $\sigma_z$ and $j_z$
are independent.
Lets us first look at Table I. The two rows correspond to different total
angular momentum $j$. 
Either of the two columns specified by $\sigma_z$ correspond also 
to the same $j_z=s_z+\sigma_z$.
Then summing up the two coefficients squared given in either of the
two rows, one finds,
\begin{eqnarray}
\sum_{\sigma_z=\pm 1/2}|\langle s,s_z;1/2,\sigma_z|j,j_z\rangle|^2&=&
\nonumber \\
\sum_{j_z=-j,\cdots,j}|\langle s,s_z;1/2,\sigma_z|j,j_z\rangle|^2
&=&{2j+1\over 2s+1}.
\label{szquad}
\end{eqnarray}
The identity (\ref{szquad}) applies to {\it a given set of $s$, $j$ and $s_z$}.
If we release the quantum number $j$ in Eq. (\ref{szquad}),
and take the summation over two spin sectors, one finds,
\begin{eqnarray}
\sum_{j=s\pm 1/2}\sum_{j_z=-j,\cdots,j}
|\langle s,s_z;1/2,\sigma_z|j,j_z\rangle|^2
\nonumber \\
={2s\over 2s+1}+{2s+2\over 2s+1}=2.
\label{CTquad}
\end{eqnarray}
which is independent of the spin $s$.
The identities (\ref{szquad}),(\ref{CTquad}) are used to obtain the expressions
for $A(s_z,s_z)$.

\begin{table}[htbp]
\begin{center} 
\begin{tabular}{c|c|c}
        & $\sigma_z=-1/2$ & $\sigma_z=1/2$ \\
\hline
$j=s-1/2$ & $\sqrt{s+s_z\over 2s+1}$ & $\sqrt{s-s_z\over 2s+1}$ \\
\hline
$j=s+1/2$ & $\sqrt{s-s_z+1\over 2s+1}$ & $\sqrt{s+s_z+1\over 2s+1}$ \\
\end{tabular} 
\end{center}
\caption{Clebsch-Gordan coefficients, 
$\langle s,s_z;1/2,\sigma_z|j,j_z=s_z+\sigma_z\rangle$
for {\it given $s$ and $s_z$}.
Note the difference between Tables I and II.
They are equivalent but represented in different ways for
different use. The four elements on the Table correspond 
one by one, i.e., they are same in number in different Tables,
but written in terms of different parameters (either $j_z$ or $s_z$).}
\end{table}
\begin{table}[htbp]
\begin{center} 
\begin{tabular}{c|c|c}
        & $\sigma_z=-1/2$ & $\sigma_z=1/2$ \\
\hline
$j=s-1/2$ & $\sqrt{s+j_z+1/2\over 2s+1}$ & $\sqrt{s-j_z+1/2\over 2s+1}$ \\
\hline
$j=s+1/2$ & $\sqrt{s-j_z+1/2\over 2s+1}$ & $\sqrt{s+j_z+1/2\over 2s+1}$ \\
\end{tabular} 
\end{center}
\caption{Clebsch-Gordan coefficients, 
$\langle s,s_z=j_z-\sigma_z;1/2,\sigma_z|j,j_z\rangle$
for {\it given $s$ and $j_z$}.
Note that for $s$ and $j_z$ given, only four elements shown in the 
Table remain finite.}
\end{table}

On the other hand, keeping $s$ and $j$ fixed, one can also consider
another situation where {\it $j_z$ is fixed} (instead of $s_z$).
This corresponds to Table II.
If one sums up the two coefficients squared given in either of the
two rows in Table II (as we did for Table I), one finds this time,
\begin{eqnarray}
\sum_{\sigma=\pm 1/2}
|\langle s,s_z;1/2,\sigma_z|j,j_z\rangle|^2&=&
\nonumber \\
\sum_{s_z=-s,\cdots,s}
|\langle s,s_z;1/2,\sigma_z|j,j_z\rangle|^2&=&1.
\label{jzquad}
\end{eqnarray}
If we focus on a particular term in the above summation,
for which $s_z$ is given, then $\sigma_z$ is determined automatically 
by the constraint $j_z=s_z+\sigma_z$.
Therefore, the summation over $\sigma_z$ and $s_z$ can be replaced by
one another.
The identity (\ref{jzquad}) is relevant for finding the diagonal elements 
of the matrix $D$ defined by Eq. (\ref{d}). It turns out that
$D_j(j_z,j_z)$ does not depend on $j_z$.
The identity (\ref{jzquad}) applies to {\it a given set of $s$, $j$ and $j_z$}.

The two identities (\ref{szquad},\ref{jzquad}) are indeed consistent,
\begin{eqnarray}
\sum_{s_z=-s,\cdots,s}\sum_{j_z=-j,\cdots,j} |\langle s,s_z;1/2,\sigma_z|j,j_z\rangle|^2=
\nonumber \\
\sum_{s_z=-s,\cdots,s} {2j+1\over 2s+1}=2j+1,
\nonumber \\
\sum_{j_z=-j,\cdots,j}\sum_{s_z=-s,\cdots,s} |\langle s,s_z;1/2,\sigma_z|j,j_z\rangle|^2=
\nonumber \\
\sum_{j_z=-j,\cdots,j} 1 =2j+1.
\label{sumquad}
\end{eqnarray}
Furthermore, releasing again  the quantum number $j$ and taking the summation
over two spin sectors, one can verify,
\begin{eqnarray}
\sum_{s_z=-s,\cdots,s}\sum_{j=s\pm 1/2}\sum_{j_z=-j,\cdots,j} 
|\langle s,s_z;1/2,\sigma_z|j,j_z\rangle|^2
\nonumber \\
=\sum_{s_z=-s,\cdots,s}2=2(2s+1),
\nonumber \\
\sum_{j=s\pm 1/2}\sum_{j_z=-j,\cdots,j}\sum_{s_z=-s,\cdots,s}
|\langle s,s_z;1/2,\sigma_z|j,j_z\rangle|^2
\nonumber \\
=\sum_{j=s\pm 1/2}(2j+1)=2(2s+1).
\nonumber
\end{eqnarray}
We used Eq. (\ref{CTquad}) for obtaining the second expression,
whereas for the third one, we used Eq. (\ref{sumquad}),
but, of course, whichever path one chooses, one always 
obtain at the end, the same $2s+1$.

\subsection{The quartic relation}

As a by-product of the calculation of FCS,
we discovered some unconventional quartic relations 
among Clebsch-Gordan coefficients.
In the evaluation of a determinant 
derived from the modified master equation (\ref{masmod})
we encountered a $(2s+1)\times (2s+1)$ square matrix, 
consisting of forth-order Clebsch-Gordan coefficients, 
defined as Eq. (\ref{xs}),
whose $(s_z,s'_z)$-elements one can rewrite here as,
\begin{eqnarray}
X_s (s_z,s'_z)=
\sum_{j=s\pm 1/2}\sum_{j_z=-j,\cdots,j}
\nonumber \\
|\langle s,s_z;1/2,\sigma_z|j,j_z\rangle|^2
|\langle s,s'_z;1/2,\sigma'_z|j,j_z\rangle|^2.
\nonumber
\end{eqnarray}
$X_s$ is a {\it tridiagonal} matrix, i.e.,
only $(s_z,s_z)$, $(s_z,s_z+1)$ and $(s_z,s_z-1)$ elements are non-zero.
They are given explicitly as,
\begin{eqnarray}
X_s(s_z,s_z)&=&
{2\{s^2+(s+1)^2+2s_z^2\}\over (2s+1)^2},
\nonumber \\
X_s(s_z,s_z+1)&=&
{2(s-s_z)(s+s_z+1)\over (2s+1)^2},
\nonumber \\
X_s(s_z,s_z-1)&=&
{2(s+s_z)(s-s_z+1)\over (2s+1)^2}.
\label{xstri}
\end{eqnarray}
This matrix $X_s$ has a very characteristic property,
Eq. (\ref{xseigen}), i.e.,
\begin{equation}
X_s \left(\begin{array}{c} 1 \\ 1 \\ \vdots \\ 1 \end{array}\right) 
=2 \left(\begin{array}{c}  1 \\ 1 \\ \vdots \\ 1 \end{array}\right),
\label{xseigen'}
\nonumber
\end{equation}
This can be checked explicitly for $s=1/2, 1, 3/2,\cdots$.
Indeed, the matrix $X_s$ reads,
\begin{eqnarray}
X_{1/2}&=&\left(\begin{array}{cc}
3/2 &1/2 \\
1/2 &3/2
\end{array}\right),
X_{1}=\left(\begin{array}{ccc}
14/9 &4/9  &0    \\
4/9  &10/9 &4/9  \\
0    &4/9  &14/9
\end{array}\right),
\nonumber \\
X_{3/2}&=&\left(\begin{array}{cccc}
13/8 &3/8 &0   &0    \\
3/8  &9/8 &1/2 &0    \\
0    &1/2 &9/8 &3/8  \\
0    &0   &3/8 &13/8 
\end{array}\right), \cdots.
\end{eqnarray}
Proving the general result (\ref{xseigen'}) for arbitrary $s$
is equivalent to establishing the following identity:
\begin{eqnarray}
\sum_{j=s\pm 1/2}\sum_{s'_z=-s,\cdots,s}\sum_{j_z=-j,\cdots,j}
\nonumber \\
|\langle s,s_z;1/2,\sigma_z|j,j_z\rangle|^2
|\langle s,s'_z;1/2,\sigma'_z|j,j_z\rangle|^2=2.
\label{sumquart}
\end{eqnarray}
When $s_z\neq \pm s$, one can verify this by substituting into it 
explicit matrix elements given in Eq. (\ref{xstri}), i.e.,
\begin{equation}
X_s(s_z,s_z-1)+X_s(s_z,s_z)+X_s(s_z,s_z+1)=2.
\label{xssum}
\end{equation}
Then one can further verify that Eq. (\ref{xssum}) still holds
when the {\it edges} of the $X_s$ matrix are encountered, at $s_z=\pm s$. 

The same kind of structure as Eqs. (\ref{xseigen'}) and (\ref{sumquart})
exists actually for each spin sector $j$.
$Y_{s,j}$ introduced in Eq. (\ref{ys}) is such a matrix,
whose $(s_z,s'_z)$-elements one can rewrite here as, 
\begin{eqnarray}
Y_{s,j} (s_z,s'_z)=
\sum_{j_z=-j,\cdots,j}
|\langle s,s_z;1/2,\sigma_z|j,j_z\rangle|^2
\nonumber \\
\times
|\langle s,s'_z;1/2,\sigma'_z|j,j_z\rangle|^2.
\nonumber
\end{eqnarray}
$Y_{s,j}$ is related $X_s$ as,
$X_s=\sum_{j=s\pm 1/2}Y_{s,j}=Y_{s,s-1/2}+Y_{s,s+1/2}$.
Again $Y_{s,j}$ is a {\it tridiagonal} matrix,
with finite matrix elements only at $(s_z,s_z)$, $(s_z,s_z+1)$ and 
$(s_z,s_z-1)$.
As expected, the matrix $Y_{s,j}$ has the following characterizing
feature, Eq. (\ref{yseigen}), i.e.,
\begin{equation}
Y_{s,j} \left(\begin{array}{c} 1 \\ 1 \\ \vdots \\ 1 \end{array}\right) 
={2j+1\over 2s+1}\left(\begin{array}{c}  1 \\ 1 \\ \vdots \\ 1 \end{array}\right),
\label{yseigen'}
\end{equation}
which is also equivalent to the following identity:
\begin{eqnarray}
\sum_{s'_z=-s,\cdots,s}\sum_{j_z=-j,\cdots,j}
|\langle s,s_z;1/2,\sigma_z|j,j_z\rangle|^2
\nonumber \\
\times
|\langle s,s'_z;1/2,\sigma'_z|j,j_z\rangle|^2=
{2j+1\over 2s+1}.
\label{quart}
\end{eqnarray}
Now in order to prove Eq. (\ref{yseigen'}) or (\ref{quart})
for general $s$ let us consider the following two cases
separately:
\begin{enumerate}
\item{$j=s-1/2$ spin sector.}
One can first directly verify Eq. (\ref{yseigen})
by writing down explicitly the matrices $Y_{s,s-1/2}$
for $s=1/2, 1, 3/2, \cdots$:
\begin{eqnarray}
Y_{1/2,0}&=&\left(\begin{array}{cc}
1/4 &1/4 \\
1/4 &1/4
\end{array}\right),
\nonumber \\
Y_{1,1/2}&=&\left(\begin{array}{ccc}
4/9 &2/9 &0    \\
2/9 &2/9 &2/9  \\
0   &2/9 &4/9
\end{array}\right),
\nonumber \\
Y_{3/2,1}&=&\left(\begin{array}{cccc}
9/16 &3/16 &0    &0    \\
3/16 &5/16 &1/4  &0    \\
0    &1/4  &5/16 &3/16 \\
0    &0    &3/16 &9/16 
\end{array}\right),\cdots.
\nonumber
\end{eqnarray}
The finite tridiagonal matrix elements are in this case,
\begin{eqnarray}
Y_{s,s-1/2}(s_z,s_z)&=&
{2(s^2+s_z^2)\over (2s+1)^2},
\nonumber \\
Y_{s,s-1/2}(s_z,s_z+1)&=&
{(s-s_z)(s+s_z+1)\over (2s+1)^2},
\nonumber \\
Y_{s,s-1/2}(s_z,s_z-1)&=&
{(s+s_z)(s-s_z+1)\over (2s+1)^2}.
\label{ys1}
\end{eqnarray}
\item{$j=s+1/2$ spin sector.} 
One verifies Eq. (\ref{yseigen})
by writing down explicitly the matrices $Y_{s,s+1/2}$
for $s=1/2, 1, 3/2, \cdots$:
\begin{eqnarray}
Y_{1/2,1}&=&\left(\begin{array}{cc}
5/4 &1/4 \\
1/4 &5/4
\end{array}\right),
\nonumber \\
Y_{1,3/2}&=&\left(\begin{array}{ccc}
10/9 &2/9 &0    \\
2/9 &8/9 &2/9  \\
0   &2/9 &10/9
\end{array}\right),
\nonumber \\
Y_{3/2,2}&=&\left(\begin{array}{cccc}
17/16 &3/16  &0     &0    \\
3/16  &13/16 &1/4   &0    \\
0     &1/4   &13/16 &3/16 \\
0     &0     &3/16  &17/16 
\end{array}\right),\cdots.
\nonumber
\end{eqnarray}
Notice also that 
$X_s=Y_{s,s-1/2}+Y_{s,s+1/2}$. 
The finite tridiagonal matrix elements are,
\begin{eqnarray}
Y_{s,s+1/2}(s_z,s_z)&=&
{2\left\{(s+1)^2+s_z^2\right\}\over (2s+1)^2},
\nonumber \\
Y_{s,s+1/2}(s_z,s_z+1)&=&
{(s-s_z)(s+s_z+1)\over (2s+1)^2},
\nonumber \\
Y_{s,s+1/2}(s_z,s_z-1)&=&
{(s+s_z)(s-s_z+1)\over (2s+1)^2}.
\label{ys2}
\end{eqnarray}
\end{enumerate}
When $s_z\neq \pm s$, one can verify Eq. (\ref{yseigen'}),
using either Eq. (\ref{ys1}) or (\ref{ys2}), i.e.,
one finds for a given $s_z$,
\begin{eqnarray}
Y_{s,j}(s_z,s_z-1)+Y_{s,j}(s_z,s_z)
\nonumber \\
+Y_{s,j}(s_z,s_z+1)={2j+1\over 2s+1}.
\label{yssum}
\end{eqnarray}
Then one can further verify that Eq. (\ref{yssum}) still holds
when the {\it edges} of the $X_s$ matrix are encountered, at $s_z=\pm s$. 
This establishes Eq. (\ref{yseigen'}), or equivalently,
(\ref{quart}).

Note that the quartic summation, Eq. (\ref{quart}),
or the eigenvalue of Eq. (\ref{yseigen'})
{\it happens to be} identical to a quadratic 
summation, (\ref{szquad}). 
This coincidence is a crucial ingredient for the FCS solvability
of our model.
If one takes the summation over two spin sectors $j=s\pm 1/2$, 
Eq. (\ref{quart}) recovers Eq. (\ref{sumquart}).

\end{appendix}




\begin{thebibliography}{99}

\bibitem{nitzan}
A. Nitzan and M.A. Ratner, Science {\bf 300}, 1384 (2003).

\bibitem{book}
G. Cuniberti, G. Fagas, and K. Richter (eds.), 
{\sl Introducing Molecular Electronics}, 
Lect. Notes Phys. 680 (Springer 2005).

\bibitem{tsukada}
M. Tsukada, K. Tagami, K. Hirose, and N. Kobayashi, J. Phys. Soc. Jpn.
{\bf 74}, 1079 (2005).


\bibitem{some_reviews}
A. Mitra, I. Aleiner and A. J. Millis,
Phys. Rev. {\bf B 69}, 245302 (2004).

\bibitem{utsumi_martinek}
Y. Utsumi, J. Martinek, G. Sch\"on, H. Imamura, S. Maekawa, Phys. Rev. {\bf B 71}, 
245116 (2005).

\bibitem{heersche} 
H. B. Heersche, Z. de Groot, J. A. Folk, H. S. J. van der Zant, 
C. Romeike, M. R. Wegewijs, L. Zobbi, D. Barreca, E. Tondello, and A. Cornia
Phys. Rev. Lett. {\bf 96}, 206801 (2006).

\bibitem{romeike_incoh} C. Romeike, M. R. Wegewijs, and H. Schoeller
Phys. Rev. Lett. {\bf 96}, 196805 (2006).

\bibitem{elste_timm}
F. Elste, C. Timm, Phys. Rev. {\bf B 71}, 155403 (2005).

\bibitem{nazarov}
Yu.V. Nazarov (ed.),
{\it Quantum noise in Mesoscopic Phyiscs},
NATO Scinece Series II: Mathematics, Physics and Chemistry, Vol. 97,
Kl\"uwer Academic Publishers, Dordrecht, 2002.

\bibitem{fazio}
R. Fazio, V.F. Gantmakher and Y. Imry (eds.),
{\it New Directions in Mesoscopic Physics (Towards Nanoscience)}, 
NATO Scinece Series II: Mathematics, Physics and Chemistry, Vol. 125,
Kl\"uwer Academic Publishers, Dordrecht, 2003.

\bibitem{levitov}
L.S. Levitov and G.B. Lesovik, 
JETP Lett. {\bf 58}, 230 (1993);
L.S. Levitov, H.-W. Lee, and G.B. Lesovik, 
J. Math. Phys. {\bf 37}, 4845 (1996);
L.S. Levitov, in Ref. \cite{fazio}, p.p. 67-91.

\bibitem{bagrets} 
D. A. Bagrets, Yu. V. Nazarov,
Phys. Rev. {\bf B 67}, 085316 (2003);
ibid., in Ref. \cite{nazarov}, p.p. 429-462.

\bibitem{utsumi_condmat}
Y. Utsumi, Phys. Rev. B, in press (cond-mat/0604082).

\bibitem{reulet}
B. Reulet, J. Senzier, and D.E. Prober,
Phys. Rev. Lett. {\bf 91}, 196601 (2003).

\bibitem{reznikov}
Yu. Bomze, G. Gershon, D. Shovkun, L. S. Levitov, M. Reznikov,
Phys. Rev. Lett. {\bf 95}, 176601 (2005).

\bibitem{kindermann}
M. Kindermann, Yu. V. Nazarov, C. W. J. Beenakker,
Phys. Rev. {\bf B 69}, 035336 (2004).

\bibitem{gustavsson_ensslin}
S. Gustavsson, R. Leturcq, B. Simovic, R. Schleser, T. Ihn, P. Studerus, K. Ensslin, D. C. Driscoll, and A. C. Gossard, Phys. Rev. Lett. {\bf 96}, 076605 (2006); 
S. Gustavsson, R. Leturcq, B. Simovic, R. Schleser, P. Studerus, T. Ihn, K. Ensslin, D. C. Driscoll, and A. C. Gossard,  Phys. Rev. B 74, 195305 (2006); 
S. Gustavsson, R. Leturcq, T. Ihn,, K. Ensslin, M. Reinwald, and W. Wegscheider, cond-mat/0607192.

\bibitem{fujisawa}
T. Fujisawa, T. Hayashi, R. Tomita, Y. Hirayama, Science {\bf 312}, 1634 (2006)

\bibitem{schoeller}
D. Boese and H. Schoeller, Europhys. Lett. {\bf 54}, 668 (2001).

\bibitem{kampen}
N.G. van Kampen, 
{\it Stochastic Processes in Physics and Chemistry},
Elsevier Science Publishers, 1992.

\bibitem{struben}
Yu.V. Nazarov, J.J.R. Struben, {\bf B 53}, 15466 (1996).

\bibitem{Akera}
H. Akera, Phys. Rev. {\bf B 60}, 10683 (1999).

\bibitem{romeike_kondo} C. Romeike, M. R. Wegewijs, W. Hofstetter, and H. Schoeller
Phys. Rev. Lett. {\bf 96}, 196601 (2006)


\end{thebibliography}
\end{document}